%% file: 0_ACL_main.tex
\pdfoutput=1
\documentclass[7.8pt]{article}
\usepackage[]{acl}
\usepackage{times}
\usepackage{latexsym}
\usepackage[T1]{fontenc}
\usepackage[utf8]{inputenc}
\usepackage{microtype}
\usepackage{inconsolata}
\usepackage{tcolorbox} 
\usepackage{tikz}
\usepackage{arydshln}
\usepackage{algpseudocode}
\algrenewcommand\textproc{\text}
\usepackage{makecell}
\usepackage{booktabs}
\usepackage{color}
\usepackage{colortbl}
\usepackage{multirow}
\usepackage{enumitem}
\usepackage{makecell}
\usepackage{threeparttable}
\usepackage{amsmath,amsfonts,mathtools}
\usepackage{pifont}

\usepackage{mathrsfs}
\usepackage{float}
\usepackage{graphicx}
\usepackage{subfigure}
\usepackage{xcolor}
\usepackage{enumitem}
\usepackage{tabularx, booktabs}

\usepackage{tcolorbox}
\definecolor{mygreen}{HTML}{00B050}

\definecolor{myorange}{HTML}{ED7D31}

\usepackage{array}
\usepackage{mathtools}
\usepackage{multirow}
\usepackage{subcaption}
\usepackage{tikz}

\definecolor{color1}{cmyk}{0.216,0.176,0,0}
\definecolor{color2}{cmyk}{0.059,0.235,0.392,0}

\newcommand{\customfootnotesize}{\fontsize{9.2pt}{9.2pt}\selectfont}
\usepackage{graphicx}
\usepackage{tikz}
\usepackage{wrapfig}
\usepackage{algorithm}
\usepackage{algpseudocode}
\algrenewcommand\textproc{\text}
\usepackage{makecell}
\usepackage{booktabs}
\usepackage{pifont}
\usepackage{multirow}
\usepackage{enumitem}
\usepackage{balance}
\usepackage{threeparttable}
\usepackage{amsmath,amsfonts,mathtools} 
\usepackage{colortbl}
\usepackage{mathrsfs}
\usepackage{float}
\usepackage{graphicx}
\usepackage{subfigure}
\usepackage{hyperref}
\usepackage{tabularx, booktabs}
\usepackage{tikz}
\usepackage{arydshln}
\usepackage{CJKutf8}

\definecolor{uc_color}{rgb}{0.99,0.24,0.63}
\definecolor{hc_color}{rgb}{0.02,0.51,0.51}
\definecolor{tc_color}{rgb}{0.99,0.55,0.09}

\tikzstyle{mybox} = [draw=black, very thick,
    rectangle, rounded corners, inner sep=10pt, inner ysep=13pt]
\tikzstyle{fancytitle} =[fill=black, text=white]

\newcommand{\M}{\textsc{Tc--Rag}}

\title{\M: \underline{T}uring--\underline{C}omplete \underline{RAG}'s Case study on Medical LLM Systems}

\author{
  \textbf{Xinke Jiang\textsuperscript{1,2}\thanks{Xinke, Yue and Rihong are equal contribution.}}, 
  \textbf{Yue Fang\textsuperscript{1,2}\footnotemark[1]}, 
  \textbf{Rihong Qiu\textsuperscript{1,2}\footnotemark[1]}, 
  \textbf{Haoyu Zhang\textsuperscript{7}}, 
  \textbf{Yongxin Xu\textsuperscript{1,2}} \\ 
  \textbf{Hao Chen\textsuperscript{8}}, 
  \textbf{Wentao Zhang\textsuperscript{9}}, 
  \textbf{Ruizhe Zhang\textsuperscript{1,2}}, 
  \textbf{Yuchen Fang\textsuperscript{10}}, 
  \textbf{Xinyu Ma\textsuperscript{1,2}}\\ 
  \textbf{Xu Chu\textsuperscript{1,2,4,6}}, 
  \textbf{Junfeng Zhao\textsuperscript{1,2,5}\thanks{Corresponding author.}}, 
  \textbf{Yasha Wang\textsuperscript{2,3,6}\textsuperscript{\textdagger}}
    \\
  \textsuperscript{1}School of Computer Science, Peking University, Beijing, China \\
  \textsuperscript{2}Key Lab of HCST (PKU), MOE; SCS, Peking University, China\\
  \textsuperscript{3}National Engineering Research Center For Software Engineering, Peking University, China \\
  \textsuperscript{4}Center on Frontiers of Computing Studies, Peking University, Beijing, China \\
  \textsuperscript{5}Big Data Technology Research Center, Nanhu Laboratory, Jiaxing, China \\
  \textsuperscript{6}Peking University Information Technology Institute, Tianjin Binhai, China \\
  \textsuperscript{7}City University of Hong Kong (Dongguan) \quad \textsuperscript{8}University of Chinese Academy of Sciences\\
  \textsuperscript{9}ShanghaiTech University \quad \textsuperscript{10}University of Electronic Science and Technology of China\\
  \small{
 \{xinkejiang, fangyue, rihongqiu\}@stu.pku.edu.cn, \{zhaojf, wangyasha\}@pku.edu.cn
  } \\
}

\begin{document}
\maketitle
\begin{abstract}
In the pursuit of enhancing domain-specific Large Language Models (LLMs), Retrieval-Augmented Generation (RAG) emerges as a promising solution to mitigate issues such as hallucinations, outdated knowledge, and limited expertise in highly specialized queries. However, existing approaches to RAG fall short by neglecting system state variables, which are crucial for ensuring adaptive control, retrieval halting, and system convergence. In this paper, we introduce the \textbf{\underline{T}}uring-- \textbf{\underline{C}}omplete--\textbf{\underline{RAG}} (\M) through rigorous proof, a novel framework that addresses these challenges by incorporating a Turing Complete System to manage state variables, thereby enabling more efficient and accurate knowledge retrieval. By leveraging a memory stack system with adaptive retrieval, reasoning, and planning capabilities, \M~not only ensures the controlled halting of retrieval processes but also mitigates the accumulation of erroneous knowledge via \texttt{\textbf{Push}} and \texttt{\textbf{Pop}} actions. In the case study of the medical and general domain, our extensive experiments on seven real-world healthcare and general-domain datasets demonstrate the superiority of \M~over existing methods in accuracy by over 7.20\%. Our code, datasets and RAG resources have been available at \url{https://github.com/Artessay/TC-RAG}
.
\end{abstract}

\input{1_Introduction}

\input{2_RelatedWork}

\input{3_Preliminaries}

\input{3_TuningComplete}

\input{4_Method}

\input{5_Experiment}

\input{6_Conclusion}


\input{8_Limitations}

\newpage

\bibliography{0_ACL_main}

\input{7_Appendix}

\end{document}

%% file: 1_Introduction.tex
\section{Introduction}

\textbf{Large Language Models (LLMs)}, such as ChatGPT~\cite{ChatGPT} and GPT-4~\cite{OpenAI2023GPT4TR}, have achieved remarkable strides in pivotal areas,
demonstrated exceptional performance across a variety of downstream tasks~\cite{kaplan2020scaling,vu2024gptvoicetasker,ma2025dressing,pmlrv235ma24a}.
In the medical domain—those medical LLMs~\cite{Wang_Liu_Xi_Qiang_Zhao_Qin_Liu_2023,Zhang_Chen_Jiang_Yu_Chen_Li_Chen_Wu_Zhang_Xiao_et,Yang_Zhao_Zhu_Zhou_Xu_Jia_Zan_2023,Zhu_Togo_Ogawa_Haseyama,Pal_Sankarasubbu}—exhibit great promise in the healthcare field, where accountability and trustworthiness are paramount~\cite{ji2023survey, song2024typing}. 
By incorporating comprehensive medical knowledge through pre-training~\cite{kaplan2020scaling}, they can support physicians in making accurate diagnoses and formulating treatment plans~\cite{Jiang_Zhang_Xu_Qiu_Fang_Wang_Tang_Ding_Chu_Zhao_et}, as well as enhance medical resource allocation~\cite{Wang_Zhao_Ouyang_Wang_Shen_Segmentor,Xu_Xu_Wang_Liu_Zhu_Mcauley_Diego,liu2024large}.
Despite medical LLMs' advancements, significant conundrums still remain, including the difficulty in avoiding factual inaccuracies (\textit{i.e.}, hallucinations)~\cite{ji2023survey,cao-etal-2020-factual,10.1145/3571730,HyKGE}, outdated knowledge~\cite{he2022rethinking}, and a lack of highly specialized expertise~\cite{kandpal2023large}.
Consequently, \textbf{R}etrieval-\textbf{A}ugmented \textbf{G}eneration (RAG)~\cite{GraphRAG,asai2023selfrag,yang2024faima,xu2025parenting,jiang2024ragraph}, which uses the external knowledge base to provide medical knowledge as contextual information to enhance content generation, combined with medical-LLMs with their massive parameterized knowledge can be likened to the expertise of doctors, is deemed as a promising and necessary solution to the aforementioned difficulties. 



However, while current approaches in enhancing LLMs with external knowledge through RAG show promise~\cite{su2024dragin,asai2023selfrag,fiare}, they have consistently overlooked the introduction of system state variables—\textbf{an essential component for ensuring adaptive control, retrieval halting, and system convergence}. Moreover, these existing RAG methods are not Turing Complete, lacking the ability to dynamically manage and monitor the retrieval process in a way that guarantees convergence to a reliable conclusion. In complex medical scenarios, where decisions often require intricate, multi-step reasoning and adaptive responses~\cite{mustafa2023building}, the absence of Turing Completeness~\cite{turing1936computable} significantly limits a system's effectiveness and reliability. This gap motivates our approach: \textbf{to construct a Turing Complete RAG System} that effectively manages state variables, using a finite logical framework to enhance the RAG process. However, how to effectively construct a Turing Complete RAG system remains unexplored and faces substantial challenges:

\noindent\textit{\textbf{\ding{182} C1. How to design a Turing Complete RAG System with the monitored state variable.}} Designing a Turing Complete RAG system requires the integration of monitored state variables that dynamically track and control the retrieval process—something that existing RAG methods lack. Current approaches~\cite{jeong2024adaptive,su2024dragin} do not have an explicit mechanism to assess whether the system has converged to a reliable conclusion, which is a critical gap. A significant challenge lies in leveraging the forward propagation of the large model to accurately compute these state variables in real-time. This involves ensuring that the state variables effectively reflect the system’s evolving context, guiding crucial decisions on whether to continue, halt, or refine the retrieval process. Managing these variables within the model’s forward pass, while maintaining adaptability to complex and varied medical queries, is essential for achieving both efficiency and accuracy, ensuring that the retrieval process finally reliably converges to an optimal conclusion.

\noindent\textit{\textbf{\ding{183} C2. \underline{Whether to}, \underline{What to}, and \underline{How to plan} retrieval for efficient and accurate knowledge to maintain optimal state.}} 
With the ability to assess the state, how to dynamically manage it to achieve the expected state is significant. 
For the real-life consultation case, doctors will decide whether to perform and what to search based on their state of mastery of this problem~\cite{cox2021diagnostic}, instead of the regardless search--which can lead to redundant information that the model already possesses, potentially causing confusion or even misleading the LLMs. Moreover, an experienced doctor can systematically analyze and plan further steps with access to a vast medical knowledge base, while a layperson might struggle in a dilemma as to choose where to start or which tools to use~\cite{react,zhu2024knowagentknowledgeaugmentedplanningllmbased}. The medical LLM is analogous to a medical expert~\cite{DISCMedLLM}, and the challenge lies in effectively utilizing the LLM's internal parameterized knowledge to retrieve to maintain an optimal state.

\noindent\textit{\textbf{\ding{184} C3. How to avoid irrelevant noise affecting system state during RAG.}} 
Since the traditional RAG's retrieval process is typically driven by query keywords~\cite{soman2023biomedical,kgrag_arxiv,Sen2023KnowledgeGL,kim2023kggpt} rather than the model's specific needs, it may introduce extensive irrelevant and noisy context. And the erroneous knowledge will continue to accumulate with the retrieval and reasoning process~\cite{react,shinn2024reflexion}, which can cause to waste token resources~\cite{HyKGE}, accumulate invalid memories, and encounter the ``lost in the middle''~\cite{liu2023lost} problems. Therefore, how to effectively remove erroneous knowledge is crucial for maintaining system state.


To address these challenges, we propose \underline{\textbf{T}}uring \underline{\textbf{C}}omplete-\underline{\textbf{RAG}}~(\textbf{\M}), a Turing Complete System for domain-specific LLMs to provide reliable and trustworthy medical analysis.
\ding{182} For \textbf{\textit{C1}}, we designed a Turing Complete RAG system with a memory stack that monitors intermediate states, ensuring the retrieval process reliably converges to an optimal conclusion.
\ding{183} For \textbf{\textit{C2}}, we extensively collected medical data and pre-trained a medical LLM, elevating its understanding from layperson to expert level, thus enhancing its reasoning and planning abilities. The model’s reasoning ability is leveraged to decide whether and what to retrieve adaptively, and its planning capacity guides tool usage and action planning, akin to how medical professionals solve complex problems.
\ding{184} For \textbf{\textit{C3}}, \M~incorporates a memory stack system with backtrack and summary operations to timely remove errors and condense redundant knowledge, mitigating accumulation of erroneous information and noise.
In summary, our contributions are as follows:
\begin{itemize}[leftmargin=*,noitemsep,topsep=2pt]
\item To the best of our knowledge, \M~is the first RAG framework to introduce the system state variable and the Turing Completeness mechanism, which could make the retrieval process controllable and halt.
\item By introducing state variable, we theoretically prove the Turing Completeness of our white-box approach.
\item \M~establishes a stack memory system, capable of adaptive retrieval, incorporating composed actions to effectively manage memory, particularly in handling harmful or noisy knowledge.
\item We open-source a meticulously curated Chinese medical pretraining dataset, along with extensive medical documents and a comprehensive knowledge graph.
\item Our thorough experimental evaluation on seven real-world Medical and general Q\&A datasets demonstrates the superior performance of \M~over existing baselines, underscoring its accuracy and explainability. Furthermore, \M~has been deployed on the online hospital platform (anonymous).
\end{itemize}

%% file: 2_RelatedWork.tex
\section{Related Work}

\subsection{Retrieval-Augmented Generation}

\textbf{Retrieval-Augmented Generation (RAG)}, introduced by \cite{lewis2020retrieval}, enhances LLM performance on knowledge-intensive tasks by integrating relevant information from external knowledge bases through prompt engineering. RAG not only mitigates hallucination issues during LLM inference but also provides up-to-date, task-specific knowledge, significantly boosting both interpretability and performance on downstream tasks~\cite{izacard2022atlas,asai2023self,asai2023retrieval}. In the biomedical field, RAG has been widely used to improve LLMs' reasoning and analytical capabilities by leveraging external medical knowledge from sources such as medical papers, textbooks, Wikipedia~\cite{jin2023retrieve,lala2023paperqa,zakka2024almanac,wang2023augmenting,xiong2024benchmarking}, and knowledge graphs~\cite{soman2023biomedical,matsumoto2024kragen}.


\textbf{Naive \& Advanced RAG}. Naive RAG typically follows a simple \textit{retrieve-and-read} approach, where relevant information is retrieved based on the initial user query, and the answer is generated using that content~\cite{soman2023biomedical,kgrag_arxiv,Sen2023KnowledgeGL,khandelwal2020generalizationmemorizationnearestneighbor,borgeaud2022improvinglanguagemodelsretrieving,ram2023incontextretrievalaugmentedlanguagemodels}. Advanced RAG, however, incorporates more sophisticated components such as retrievers~\cite{qu2021rocketqa, ma2023chainofskills}, rerankers~\cite{cheng2021unitedqa,yu2022kgfid}, filters~\cite{HyKGE}, and readers~\cite{yoran2024makingretrievalaugmentedlanguagemodels, fang2024enhancingnoiserobustnessretrievalaugmented}, to improve the quality of both retrieval and generation.
However, neither naive nor advanced RAG considers whether the LLM already possesses the necessary knowledge. This often leads to the retrieval of excessive, redundant information, which can mislead the model and cause a “lost in the middle” dilemma~\cite{liu2023lost}. Our method addresses this by determining whether to retrieve and what to retrieve based on the model's internal parameterized knowledge, resulting in more efficient and accurate retrieval.

\textbf{Adaptive RAG}. 
Recent research has focused on developing adaptive RAG strategies, enabling LLMs to determine whether and when to retrieve and to select the most appropriate retrieval tools from huge knowledge base. 
FLARE~\cite{jiang2023active} predicts the next sentence and uses the generated low-confidence tokens as query to re-retrieve relevant documents. 
DRAGIN~\cite{su2024dragin} leverages the LLM's uncertainty in its generated content to decide when to trigger retrieval based on the internal self-attention weights and corresponding keywords.
Adaptive-RAG~\cite{jeong2024adaptive} uses a smaller LLM as a classifier to query complexity and subsequently selects the most appropriate retrieval strategy—ranging from simple to advanced.
However, these existing adaptive RAG methods are not Turing Complete, lacking the ability to dynamically manage and monitor the retrieval process in a way that guarantees convergence to a reliable conclusion. Moreover, they have yet to fully harness the step-by-step planning and tool-use abilities of LLMs in conjunction with RAG. Our approach addresses these limitations by integrating a Turing Complete framework that optimizes retrieval through advanced planning and tool-use strategies, ensuring more reliable and accurate outcomes.

\subsection{Reasoning and Planning Capabilities}

Recent advancements have focused on enhancing the reasoning and planning capabilities of LLMs~\cite{zhu2024knowagentknowledgeaugmentedplanningllmbased,react,shinn2024reflexion,koh2024treesearchlanguagemodel}. One notable approach is Chain-of-Thought (CoT)~\cite{wei2022chain,xiachain}, which demonstrates how LLMs can construct structured "thought processes" to solve complex problems. ReAct\cite{yao2022react} integrates reasoning traces with task-specific actions, enabling LLMs to plan, adjust actions, and manage exceptions while gathering information from external sources such as knowledge bases. Reflexion~\cite{shinn2024reflexion} further improves LLMs by using linguistic feedback, allowing them to reflect on and store task feedback, enhancing decision-making in future trials. Despite these advancements, the reasoning and planning processes of LLMs often lead to the accumulation of errors and redundant information. While these methods introduce new decision trials, they frequently fall short in managing previous memory—particularly in removing ineffective decisions or refining historical records.
To address these challenges, \M~incorporates a memory stack system with backtrack and summary operations, allowing for timely error correction and condensation of redundant knowledge. This ensures that the model’s reasoning process remains efficient and accurate, leading to more reliable outcomes.

%% file: 3_Preliminaries.tex
\section{Prelinmary}

\paragraph{\textbf{Definition 1. (Stack of Memory).}}
\label{definition1}
In \M, the Memory of LLMs is conceptualized as a White Box Turing Machine~\cite{turing1936computable}, with the rigorous theoretical proof in \textbf{Appendix~\ref{Turing-Complete}}.
Let $\text{\textsc{Tc}} = (\mathcal{S}, \mathcal{A}, \mathcal{M}, \delta, s_0, \mathcal{F}, \sigma)$ represent the stack memory of the LLMs, where:
\begin{itemize}[leftmargin=*,noitemsep,topsep=2pt]
    \item $\mathcal{S}$ denotes the set of possible states the LLMs occupy. Specifically, we use numerical values like perplexity~\cite{jelinek1977perplexity} \& uncertainty~\cite{peng2024revisitingdemonstrationselectionstrategies} to define system's state.

    \item $\mathcal{A}$ represents the set of actions that the LLMs can perform. Following stack theory, we define the fundamental meta-actions: \textbf{\texttt{push}} and \textbf{\texttt{pop}}. Additionally, we have composite actions composed of the two meta-actions: ``\texttt{Thought}'', ``\texttt{Tool\_Observation}'', ``\texttt{Plan}'', ``\texttt{Backtrack}'', ``\texttt{Summary}'', and ``\texttt{Conclusion}''.

    \item $\mathcal{M}$ is the infinite stack memory, which includes the set of actions $\mathcal{A}$ and potentially other symbols may be used in stack operations, such as the initiation state ``\texttt{User\_Query}''.

    \item $\delta : \mathcal{S} \times \mathcal{M} \rightarrow \mathcal{S} \times \mathcal{A} \times \mathcal{M} \times (0, +\infty)$ is the state transition function. It defines how the LLM transitions to a new state, selects an action, updates the stack memory, and calculates a new system state value based on the current state and stack memory. The transition involves selecting an action $a \in \mathcal{A}$ (which includes push and pop operations), leading to a new stack top and an updated system state value $s \in \mathcal{S}$ within the range $(0, \text{\$Large\_Value\$})$. $\delta$ is decided by LLMs and state calculation function.

    \item $s_0 \in \mathcal{S}$ is the initial state, with its initial and upper bound value defined as $s_0 = \text{\$Large\_Value\$} $ (where a lower value indicates a more desirable system state).

    \item $\mathcal{F} \subseteq \mathcal{S}$ is the set of final states. The state $f \in \mathcal{F}$ is deemed final if the current action is ``\texttt{Conclusion}'' and the current state value $f$ is less than the threshold $\sigma$. Otherwise, the system will either reset the state value to the threshold $\sigma$ or change the action from ``\texttt{Conclusion}'' to ``\texttt{Thought}'' for further analysis.

    \item $\sigma \in (0,+\infty)$ is the hyper-threshold for the final state.
\end{itemize}

\paragraph{\textbf{Definition 2. (Meta-Actions).}}
In this part, we define the meta-actions in \M~as follows:
\begin{itemize}[leftmargin=*]
\item \texttt{Push}: When the medical-LLM processes new information or obtains tool observations, a new action is determined and pushed onto the stack $\mathcal{M}$:
\noindent
\begin{footnotesize}
\begin{align*}
    (s_{t+1}, a_{t+1}) \leftarrow \delta(s_t, \mathcal{M}_t), \mathcal{M}_{t+1} \leftarrow  push(\mathcal{M}_t, a_{t+1}), 
\end{align*}
\end{footnotesize}
where $s_{t+1}$ represents LLM's next system state value, $a_{t+1}$ is the next action to be pushed onto stack $\mathcal{M}$. 

\item \texttt{Pop}: If an error or inconsistency is detected, the LLM needs to revert to the previous state by popping the top element from the stack:
\begin{footnotesize}
\begin{align*}
    (s_{t+1}, a_{t+1}) \leftarrow \delta(s_t, \mathcal{M}_{t}), \quad \mathcal{M}_{t+1} \leftarrow pop(\mathcal{M}_t).
\end{align*}
\end{footnotesize}
\end{itemize}

\paragraph{\textbf{Task Definition.}}
Given the parameterized knowledge $\Theta$ embedded within LLM, and non-parameterized knowledge base $\mathcal{D}$ which comprises diverse types of knowledge,
the objective is to generate a reliable medical \texttt{Response} given the natural language \texttt{User\_Query}: 
\begin{equation}
\texttt{Response} \leftarrow \Theta(\texttt{User\_Query}, \mathcal{D}~|~\mathcal{P}),
    \label{eq:prediction objective}
\end{equation}
where $\mathcal{P}$ is the task-specific prompt, \textit{i.e.} CoT's~\cite{wei2023chainofthought}, ReACT's~\cite{react}, Trigger Prompt's~\cite{xu2025dearllm}, etc.

\paragraph{\textbf{Reasons for Stack Memory.}}  
The stack design in \M~offers several advantages: \ding{182} ensures the purity of historical information by removing irrelevant knowledge via \texttt{pop}; \ding{183} provides $O(1)$ access efficiency for fast retrieval of past information; \ding{184} memorizes both parameterized and external knowledge, enhancing reasoning capabilities; \ding{185} allows dynamic control of the reasoning process with options for backtracking and summarizing; \ding{186} and mimics human-like reasoning, where information is accumulated and conclusions are adjusted as new evidence arises. (cf \textbf{Appendix}~\ref{reasons for stack memory} for details).

%% file: 3_TuningComplete.tex
\newtheorem{theorem}{Theorem}
\newtheorem{lemma}{Lemma}
\newtheorem{proof}{Proof}[section]

\section{TURING-COMPLETE}
\label{sec: main Proof of TURING-COMPLETE}
We define a stack-based memory system incorporating a system state variable and prove its equivalence to a universal Turing machine $T$ through a series of formal definitions, lemmas, and a main theorem in 5 steps. The detailed proofs are provided in \textbf{Appendix~\ref{Turing-Complete},~\ref{appendix:convergence}}.
We prove this by showing that for any Turing machine $T$, there exists \textsc{Tc} that can simulate $T$.  

\paragraph{Step \ding{182}: Construct Turing System}
For a Turing machine $T = (\mathcal{S}_T, \mathcal{A}_T, \mathcal{M}_T, \delta_T, st_0, st_{accept}, st_{reject})$, we interpret $\text{\textsc{Tc}} = (\mathcal{S}, \mathcal{A}, \mathcal{M}, \delta, s_0, \mathcal{F}, \sigma)$ with the following components:

\begin{enumerate}[leftmargin=*,noitemsep,topsep=2pt]
\item $\mathcal{S} = \mathcal{S}_T \cup {f}$, where $f \notin \mathcal{S}_T$ is the end state.
\item $\mathcal{A} = \mathcal{A}_T$, $\mathcal{M} = \mathcal{M}_T \cup \mathcal{A}$, $s_0 = st_0$, $\mathcal{F} = \{f\}$, $\sigma$ is the hyper-threshold to decide accept or reject.
\item The transition function $\delta$ is defined as:
\end{enumerate}
\begin{footnotesize}
\begin{align*}
\delta(s,a)=\begin{cases}
        (s',push,s\geq\sigma)&\text{if } \delta_T(st, a)=(st',a,R) \\
        (s',pop,s\geq\sigma)&\text{if } \delta_T(st,a)=(st',a,L) \\
        (f,no\_op,a, s<\sigma)&\text{if } st\in\{st_{accept},st_{reject}\}
    \end{cases}
\end{align*}
\end{footnotesize}
where $a$ is the executing action for $T$ or \textsc{Tc}. The operations $push$ and $pop$ are corresponding to the right ($R$) and left ($
L$) movements in $T$. Transition function $\delta(\cdot,\cdot)$ is for \textsc{Tc} and $\delta_T(\cdot,\cdot)$ is for $T$. $no\_op$ indicates no operation is performed when the system is halting.


\paragraph{Step \ding{183}: Configuration Mapping}
Next, we define a bijective mapping function $h$ that maps each configuration of $T$ to \textsc{Tc}. 
We now define configuration $c$ of $T$ as $c=(st, w_1 a w_2)$, $w_1, w_2$ represents the sequence of actions to the left / right of the tape head, where $st \in \mathcal{S}_T$, $a \in \mathcal{A}_T$ is the tape head/action.
Configuration of \textsc{Tc} $c_{\textsc{Tc}
}$ is composed of the tuple as (system state, remaining actions, stack memory actions, whether to halt) as:
\[ h(st, w_1aw_2) = (s, w_2, w_1^Ra, s\geq \sigma) = c_{\textsc{Tc}
}\]
where $w_1^R$ is the reverse of $w_1$, which is necessary due to Last-In-First-Out (LIFO) principle of $\mathcal{M}$.

\paragraph{Step \ding{184}: Simulation of Computation Steps}
We prove that \textsc{Tc} can simulate each step of $T$ (Lemma~\ref{lemma 1}):
\begin{tcolorbox}[colback=black!5!white, colframe=black!50!white, 
                  sharp corners, boxsep=0pt, left=2pt, right=2pt, top=2pt, bottom=2pt]
\begin{lemma}
\label{lemma 1}
If $c_1\vdash_T c_2$ in $T$, then $h(c_1)\vdash_{\textsc{Tc}}^* h(c_2)$ in \textsc{Tc}, where $*$ denotes one or multiple steps of derivation, $\vdash$ is the action shift operator.
\end{lemma}
\end{tcolorbox}

\begin{proof}
Let $c_1 = (st_1, w_1a_1w_2)$ and $c_2 = (st_2, w_1'a_2w_2')$ be configurations of $T$ with $a_t=\{L,R\}$. We prove Lemma~\ref{lemma 1} by considering the two cases $push$ and $pop$ corresponding to the possible directions of tape head movement in $T$.
\begin{itemize}
[leftmargin=*,noitemsep,topsep=2pt]
    \item \textbf{Case 1}: If $\delta_T(st_1, a_1) = (st_2, a_2, R)$, then:
    \begin{align*}
    h(c_1) &= (st_1, w_2, w_1^Ra_1, st_1\geq\sigma) \\
    &\vdash_{\textsc{Tc}} (s_2, {w_2'}, w_1^Ra_1a_2, s_2\geq\sigma) \quad (\text{\textbf{push} $a_2$})\\
    &= h(st_2, \underline{w_1a_1}{a_2w_2'})= h(st_2, \underline{w_1'}a_2w_2')\\&= 
    h(c_2)
    \end{align*}
    
    \item \textbf{Case 2}: If $\delta_T(st_1, a_1) = (st_2, a_2, L)$, then:
    \begin{align*}
    h(c_1) &= (st_1, w_2, w_1^Ra_1, st_1\geq\sigma) \\
    &\vdash_{\textsc{Tc}} (s_2, w_2, w_1^R, s_2\geq\sigma) \quad (\text{\textbf{pop} $a_1$}) \\
    &\vdash_{\textsc{Tc}} (s_2, w_2', w_1'^R a_2, s_2\geq\sigma) \\&  \quad \text{(\textbf{push} $a_2$, if $a_2 \neq \text{null}$)} \\
    &= h(st_2, w_1'a_2w_2') = h(c_2)
    \end{align*}
\end{itemize}
\end{proof}
It's noted that the two cases correspond to meta operations of \texttt{push} (Case 1) and \texttt{pop} (Case 2) in \M.

\paragraph{Step \ding{185}: Preservation of Acceptance and Rejection}
\begin{tcolorbox}
[colback=black!5!white, colframe=black!50!white, 
                  sharp corners, boxsep=0pt, left=2.2pt, right=2.2pt, top=3pt, bottom=2.2pt]
\begin{lemma}
$T$ accepts (rejects) whole input $w$ if and only if \textsc{Tc} reaches a configuration $(f, NULL, \mathcal{M}, f<\sigma)$ from the initial configuration $(s_0, w, NULL, s_0\geq \sigma)$.
\end{lemma}
\end{tcolorbox}
\begin{proof}
We prove the consistency of reaching the termination state from both $T\to \textsc{Tc}$ and $\textsc{Tc} \to T$ perspectives.

(\textbf{Forward} $\Rightarrow$) Assume $T$ accepts (rejects) input $w$. Then:
$\exists t \in \mathbb{N}$ such that after $t$ steps, $T$ enters state $st_{accept}$ (or $st_{reject}$). Then, Let $(st_t, w_t)$ be the configuration of $T$ at step $t$, where $st_t \in \{st_{accept}, st_{reject}\}$. By our construction of $\delta$, $\forall a \in \mathcal{A}$, for $\forall st \in \{st_{accept}, st_{reject}\}$ we have:
    \[\delta(s, a) = (f, no\_op, a, f<\sigma)\]
Therefore, if $h(st_t, w_t) = (s_t, w_2, w_1^Ra, s_t \geq \sigma)$ is the corresponding configuration in \textsc{Tc}, then:
    \[(s_t, w_2, w_1^Ra, s_t\geq \sigma) \vdash_{\textsc{Tc}}^t (f, NULL, \mathcal{M}, f<\sigma)\]
Thus, \textsc{Tc} achieves $f$ with state value lower than $\sigma$.

(\textbf{Backward} $\Leftarrow$) Assume \textsc{Tc} reaches configuration $(f, NULL, \mathcal{M}, f<\sigma)$ from $(s_0, w, NULL, s_0\geq \sigma)$. Then:
$\exists t \in \mathbb{N}$ such that:
\[(s_0, w, NULL, s_0\geq \sigma) \vdash_{\textsc{Tc}}^t (f, NULL, \mathcal{M}, f<\sigma).\] 
Let $(s_{t-1}, w_2, w_1^Ra, s_{t-1} \geq \sigma)$ be the configuration of \textsc{Tc} at step $t-1$. Then, by the construction of $\delta$, the only way to reach $f$ is if $s_{t-1} \in \{st_{accept}, st_{reject}\}$.
Therefore, the corresponding configuration of $T$ at step $t-1$ must be $(st_{t-1}, w_1aw_2)$, where $st_{t-1} \in \{st_{accept}, st_{reject}\}$.
Thus, $T$ must have entered $st_{accept}$ or $st_{reject}$, implying that $T$ accepts or rejects $w$.

From both the \textbf{Forward} and \textbf{Backward} proofs, $T$ accepts (or rejects) the input $w$ if and only if \textsc{TC} reaches a configuration $(f, \text{NULL}, \mathcal{M}, f < \sigma)$ from the initial configuration $(s_0, w, \text{NULL}, s_0 \geq \sigma)$.
\end{proof}

\paragraph{Step \ding{186}: Convergence of the State Value to \( \sigma \)}
As detailed in Appendix~\ref{appendix:convergence}, under the \textit{Action Validity Assumption}, we have shown that the Lyapunov function \( V(st_j) = \max(0, st_j - \sigma) \) satisfies the following key properties:
\(
V(st_{j+1}) - V(st_j) \leq -\epsilon,  \epsilon > 0,
\)
which ensures that the system state \( s_t \) decreases monotonically. As a result, the system state will eventually reach the threshold \( \sigma \), satisfying \( st_T < \sigma \) for some finite \( T \). This guarantees finite-time convergence. And the convergence time \( T \) satisfies the following bound with approximate     $
\frac{\log(\delta / V(s_{t_0}))}{\log(1 - \beta)}.
$
Thus, the system halts reliably after a finite number of steps.

By the above lemmas, we show that: \ding{182} \textsc{Tc} can simulate every move of $T$. \ding{183} \textsc{Tc} halts if and only if $T$ achieve acceptance or rejection behavior. \textbf{Thus, \textsc{Tc} can simulate any computation of any Turing machine $T$, which by definition makes \textsc{Tc} Turing complete}. Theoretically, \textsc{Tc} can be considered a \textbf{Tape Model}, however, considering the issue of accumulated erroneous and disordered context, in implementation, we allow the observation to the entire memory stack $\mathcal{S}$ (not just the top), but restrict LLMs to modify any position of $\mathcal{S}$. 



%% file: 4_Method.tex
\begin{figure*}[t]
  \centering
\includegraphics[width=1.02\textwidth]{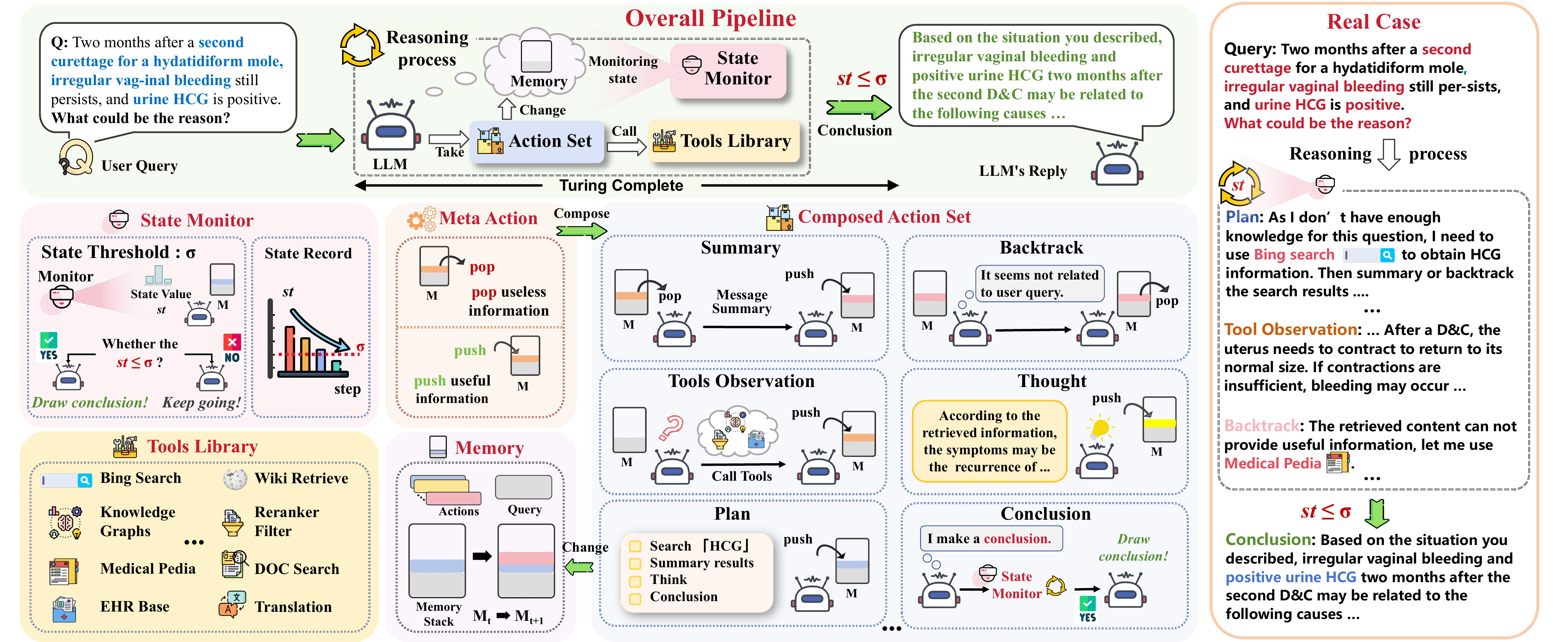}
  \caption{Overall framework of \M. 
  }
  \label{fig:newmethod.png}
\end{figure*}

\section{Memory Stack and State Monitor}
In this section, we present our innovative framework, \M, as illustrated in Figure~\ref{fig:newmethod.png}. Traditional methods, such as those based on thought chain~\cite{con, cok, ma2023chainofskills, wei2023chainofthought} and reasoning and acting approaches ~\cite{react,shinn2024reflexion,zhu2024knowagentknowledgeaugmentedplanningllmbased}, often suffer from lack of state management,  retrieval halting, accumulated erroneous knowledge, token inefficiency, and the issue of being ``lost in the middle''. To address these challenges, we introduce a stack-based memory system that leverages \texttt{push} and \texttt{pop} actions for efficient memory management for \textbf{\textit{C1, C3}}.
Additionally, to further improve the LLMs' knowledge comprehension and adaptability, we incorporate an expert knowledge pre-training module based on the general LLM, which enhances the LLMs' understanding and reasoning capabilities (as detailed in \textbf{Appendix~\ref{sec:model pretraining}}) for \textbf{\textit{C2}}. Next, in subsection~\ref{sec: memory definite and meta actions}, we define the stack memory structure and the composite actions such as ``\texttt{Thought}'', ``\texttt{Tool\_Observation}'',  ``\texttt{Plan}'', ``\texttt{Backtrack}'', ``\texttt{Summary}'', and ``\texttt{Conclusion}''. 
We then outline the stack states—initiation, intermediate, and final—and provide specific state calculations in subsection~\ref{sec: start and end}.
In \textbf{Appendix~\ref{sec: prompts and algorithm}}, we describe the prompting strategy and the full algorithm. 

\subsection{Stack Memory and Composed Actions}
\label{sec: memory definite and meta actions}
In managing memory for LLMs, memory can be conceptualized as a stack structure adhering to the ``last-in, first-out'' (LIFO) principle. 
Unlike Markov processes where future actions depend solely on the current actions, in \M, each new memory entry builds upon the accumulated memory~\cite{pimentel2018review,miller2003memory}. 
To maintain the orderliness and consistency of the memory, erroneous entries are ejected sequentially from the top of the stack. This approach prevents ``contamination'' and confusion that might arise from directly removing intermediate memories, thereby mitigating unreasonable interference during the reasoning process.

Consequently, as outlined in Definition 1, we leverage the stack $\mathcal{M}$ to simulate the LLM's memory.
Based on meta stack operations of push \& pop in Section~\ref{sec: main Proof of TURING-COMPLETE}, the following composite actions have been devised to enhance reasoning and planning capabilities of LLMs:
\begin{itemize}[leftmargin=*,noitemsep,topsep=2pt]
    \item \texttt{Thought}: This action represents LLM's analytical process. Once the model has processed and generated an insight, this analysis is \textbf{pushed} onto stack $\mathcal{M}$.
    \item \texttt{Tool\_Observation}: When the LLM interacts with external tools to gather additional data or insights, the result of this interaction is captured as an observation. Both the tool's name and the resulting observation are then \textbf{pushed} onto the stack $\mathcal{M}$.
    \item \texttt{Plan}: Triggered when the LLM identifies a relevant or helpful task, this action occurs when the model makes a decision to outline subsequent actions or strategies. The model \textbf{pushed} a new element to the stack $\mathcal{M}$, representing a planned step, decision, or strategy that will guide the next phase of the task.
    \item \texttt{Backtrack}: Triggered by the LLM's reflection mechanism, this action occurs when the top of the stack is found to be irrelevant or harmful to the ongoing task. The model \textbf{pops} the top element off the stack $\mathcal{M}$, removing the irrelevant or harmful memory.
    \item \texttt{Summary}:
    When the content at the top of the stack $\mathcal{M}$ becomes too lengthy or cluttered with irrelevant information, the LLM initiates a summarizing process. The content is first \textbf{popped} off the stack, and then the concise summary is \textbf{pushed} back onto the $\mathcal{M}$, ensuring that the stack remains focused on the most pertinent information.
    \item \texttt{Conclusion}: 
    This action occurs when the LLM believes it has reached a final conclusion or solution in a black-box manner. However, if the current state does not satisfy the necessary stopping conditions, the \texttt{Conclusion} action is replaced with \texttt{Thought}, allowing the model to continue refining its reasoning.
\end{itemize}

\subsection{Stack Status and Memory Management}
\label{sec: start and end}
\subsubsection{Pipeline} We will systematically describe the changes in the memory stack and system status.
\begin{itemize}[leftmargin=*,noitemsep,topsep=2pt]
    \item \textbf{Stack and State Initialization}: Initially, the overall system state value is assigned a large constant as \$Large\_Value\$. The memory stack, $\mathcal{M}$, begins with only the 
    ``\texttt{User\_Query}'' action.
    This entry is immutable, meaning it cannot be popped and permanently resides at the bottom of the stack.
    \item \textbf{State Value Evaluation}: During analytical process, outputs tagged with actions ``\texttt{Conclusion}'' and ``\texttt{Thought}'' are monitored. For each output, a system state value is computed and carried forward in subsequent iterations. This value can be determined using metrics like conditional perplexity and uncertainty, with lower values indicating higher confidence.
    \item \textbf{Conclusion Validation}: If the LLMs output a \texttt{Conclusion}, the system evaluates whether the associated state value meets predefined stopping conditions.
    If not, action \texttt{Conclusion} is reclassified as \texttt{Thought}, signaling the need for further analysis.
    \item \textbf{State Restoration}: 
    If a \texttt{Thought} is popped from the stack (\textit{e.g.}, via \texttt{Backtrack} action), the system restores the system state value to its previous level before the addition of that \texttt{Thought}. This management strategy ensures the system remains dynamic and responsive to memory, allowing for iterative refinement of LLM's outputs. By continuously evaluating and adjusting based on system state value, LLMs can effectively navigate toward accurate and reliable conclusions.
\end{itemize}

\subsubsection{System Value Calculation}
To ensure that the system is progressing toward a reliable conclusion, a system state value is calculated by comparing the content generated by the LLMs at the top of the stack (from either a \texttt{Thought} or \texttt{Conclusion}) with the bottom ``\texttt{User\_Query}''. This comparison, based on metrics such as conditional perplexity and uncertainty, quantifies how closely the system's current reasoning aligns with the original intent of the task:
\begin{itemize}[leftmargin=*]
    \item \textbf{Conditional Perplexity:}
    Conditional Perplexity is a metric used to evaluate the predictability of a language model given prior context~\cite{jelinek1977perplexity}. For the sequence of text $\mathcal{M}_{\text{top}}$ at the top of the $\mathcal{M}$, and the original query at the bottom (denoted as $\mathcal{M}_{\text{bottom}}$), the conditional perplexity $cppl$ is calculated as:    
    \begin{equation}
    \begin{aligned} &cppl(\mathcal{M_{\text{top}}}\mid\mathcal{M}_{\text{bottom}}) 
    = \exp\Bigl(-\frac{1}{N} \sum_{i=1}^{N} \log \\&\quad P({w}_i \mid{w}_1, \dots, {w}_{i-1}, \mathcal{M}_{\text{bottom}})\Bigl),
    \end{aligned}
    \end{equation}
    where $N$ is the number of tokens in $\mathcal{M}_{\text{top}}$ and $P({w}_i \mid {w}_1, \dots, {w}_{i-1}, \mathcal{M}_{\text{bottom}})$ represents the probability of the $i$-th token $w_i$ in $\mathcal{M}_{\text{top}}$ given the preceding tokens and the context $\mathcal{M}_{\text{bottom}}$.
    \item \textbf{Uncertainty:} $uct$ measures the confidence~\cite{Jiang_2023} in the context $\mathcal{M}_{\text{top}}$ and is derived from the entropy of the output probability distribution:
    \begin{equation}
    \begin{aligned}
    &{uct}(\mathcal{M}_{\text{top}}) 
    = -\sum_{i=1}^{N} P(w_i \mid {w}_1, \dots,{w}_{i-1})\\ & \quad \times \log P(w_i \mid {w}_1, \dots, {w}_{i-1}),
    \end{aligned}
    \end{equation}
    where higher entropy indicates greater uncertainty and less reliability.
\end{itemize}
The system will then select either $cppl$ or $uct$ as the system state value $s_t$ at step $t$. If $s_t < \sigma$, it suggests that the LLM's output is both predictable and confident, indicating alignment with the original query. Vice versa, a high-value signal that the output may not be reliable or relevant, necessitating further refinement.
The system will continuously update $s_t$ during reasoning process, using it to guide decisions on whether to finalize or continue refining the output.

\subsection{Acceleration Strategy}
To mitigate the time overhead associated with multi-turn interactions with LLMs, we employ two acceleration strategies:
\ding{182} \textbf{Retrieval and Interaction in the Decoder Stage.} As multi-turn interactions accumulate more contextual information, the increasing length of the context results in repeated encoding of historical context (Key and Value matrices) during the decoding process leading to significant inefficiencies. To address this, we incorporate a retrieval and interaction mechanism during the decoder stage, which interrupts the decoding process after each action is generated by LLMs~\cite{RICHES, asai2023selfrag}.
\ding{183} \textbf{Token Speculative Sampling.} The decoding stage in LLMs is computationally intensive due to the token-by-token generation process. To alleviate this, we fine-tune a smaller language model (with only 1.5B parameters) as a draft model and employ speculative sampling to predict the next token~\cite{leviathan2023fastinferencetransformersspeculative,chen2023acceleratinglargelanguagemodel}.

%% file: 5_Experiment.tex
\section{Experiments}

In this section, we conduct a series of experiments on three medical datasets to answer the following research questions:
\begin{itemize}[leftmargin=*]
    \item \textbf{RQ1} (\textbf{Section}~\ref{Main Results}, \textbf{Appendix}~\ref{appendix:Generalization Experiments}): Does \M~outperform SOTA RAG methods under the same database source?
    \item \textbf{RQ2} (\textbf{Section} \ref{Ablation Study}, \ref{EffiencyStudy}): Is the stack framework we designed effective? What impact does each component have on the overall performance?
    \item \textbf{RQ3} (\textbf{Appendix}~\ref{Noise Poisoning Attack}, \ref{Case Study}): Can \M~really pop up erroneous execution memory and noise injection and achieve memory management?
\end{itemize}
Moreover, we also anonymously rank \M~on the largest Chinese medical CMB Leaderboard, which includes detailed medical scores for each subcategory (c.f. \textbf{Appendix~\ref{Leaderboard}})\footnote{https://cmedbenchmark.llmzoo.com/static/leaderboard.html}.

\subsection{Experimental Setup}
\textbf{\ding{182} Datasets.} Our experiments are conducted on two multi-task medical datasets MMCU-Medical~\cite{mmcu} and CMB-Exam~\cite{cmb}, and one open-domain Q\&A medical dataset CMB-Clin~\cite{cmb}. \textbf{\ding{183} RAG Tools.} We incorporate various RAG tools including knowledge graph search, document search, web and pedia search, and electronic medical record database. \textbf{\ding{184} LLM Turbo.} The general-domain LLM Qwen-32B~\cite{yang2024qwen2technicalreport} was selected as the base model, which we further pre-trained for the medical domain.\textbf{\ding{185} Baselines.} To assess the effectiveness of our approach, we compare \M~with twelve baselines. (1) Without RAG: \textbf{Base} and \textbf{CoT}~\cite{wei2023chainofthought}. (2) Naive RAG: \textbf{SR-RAG}~\cite{soman2023biomedical}, \textbf{FL-RAG}~\cite{khandelwal2020generalizationmemorizationnearestneighbor} and \textbf{FS-RAG}~\cite{trivedi2023interleavingretrievalchainofthoughtreasoning}. (3) Advanced RAG: \textbf{CoK}~\cite{li2023chainofknowledge}, \textbf{SuRe}~\cite{chen2024improving}, \textbf{HyKGE}~\cite{HyKGE}. (4) Adaptive RAG: \textbf{ReACT}~\cite{react}, \textbf{Self-RAG}~\cite{asai2023selfrag}, \textbf{FLARE}~\cite{fiare} and \textbf{DRAGIN}~\cite{su2024dragin}. \textbf{\ding{186} Evaluation Metrics.}
We use \textbf{EM}~\cite{f1,f2} metric for multi-task medical choice questions and \textbf{ROUGE-R}~\cite{xu2023contextaware} and \textbf{BLEU-1, BLEU-4}~\cite{xu2023contextaware} for open-domain Q\&A tasks.
Detailed experimental settings are in \textbf{Appendix~\ref{Detailed Experimental Setup}}.

\begin{table*}[htbp]
\setlength{\tabcolsep}{0.85mm}
\caption{Performance comparison (\%) on \textit{CMB}, \textit{MMCU} and \textit{CMB-Clin} datasets.
}
\label{tab:comparison}
\centering
\resizebox{\linewidth}{!}
{
\begin{tabular}{ccc|ccccc|ccccc}
\hline
\multirow{3}{*}{\textbf{Method}} & \multicolumn{2}{c|}{\textbf{LLM Turbo}} & \multicolumn{5}{c|}{\textbf{Qwen-32B}} & \multicolumn{5}{c}{\textbf{Pretrained Qwen-32B}} \\
\cline{2-13} 
& \multirow{2}{*}{\textbf{Type}} & \textbf{Dataset} & CMB & MMCU & \multicolumn{3}{c|}{{CMB-Clin}} & CMB & MMCU & \multicolumn{3}{c}{{CMB-Clin}} \\
&  & \textbf{Metric} & EM & EM & BLEU-1 & BLEU-4 & ROUGE & EM & EM & BLEU-1 & BLEU-4 & ROUGE \\ 
\hline
\multirow{12}{*}{\textbf{Baselines}} 
& \multirow{2}{*}{Without RAG} & Base & 67.33 & 80.92 & 06.84 & 10.82 & 24.10 & 68.34 & 80.70 & 06.93 & 11.24 & 23.75 \\
& & CoT & 70.88 & 80.53 & 07.23 & 11.64 & 24.30 & 69.94 & 80.06 & 07.37 & 11.67 & 23.54 \\
\cdashline{2-13}
& \multirow{3}{*}{Naive RAG} & SR-RAG & 66.73 & 80.31 & 09.48 & 14.18 & 22.23 & 68.54 & 83.15 & 11.56 & 16.72 & 22.38 \\
& & FL-RAG & 67.94 & 75.17 & 10.67 & 15.32& 22.96 & 69.48 & 78.61 & 10.95 & 16.20 & 23.13 \\
& & FS-RAG & 64.65 & 71.62 & 05.12& 08.63 & 20.29 & 67.20 & 75.20 & 05.76 & 09.20 & 21.34 \\
\cdashline{2-13} 
& \multirow{3}{*}{Advanced RAG} & CoK & 76.25   & 79.04 & 13.96 & 25.33 & 31.49 & 79.18 & 81.31 & 14.92 & 27.84 & 46.90\\
& & SuRe & 76.96 & 83.08 & 14.88 & 27.90 & 31.06 & 78.55 & 85.56 & 15.26& 28.01& 44.38 \\
& & HyKGE & 82.17 & 86.45 & 19.02 & 41.84 & 38.92 & 84.58 & 87.09  & 22.85  & 43.79 & 50.23\\
\cdashline{2-13}
& \multirow{4}{*}{Adaptive RAG} &ReACT & 82.20  & 87.34 &19.15 & 48.55 & 41.22 & 84.08 & 88.19 & 22.37 & 49.96 & 50.45\\
& & Self-RAG & 81.88 & 85.78 & 18.23 & 46.82 & 44.70 & 84.32 & 86.77 & 20.06 & 47.12 & 48.90 \\
& & FLARE & 82.89 & 87.26 & 19.85 & 49.19 & 51.15 & 85.14 & 87.37 & 20.91 & 50.38 & 52.96 \\
& & DRAGIN & 82.78 & 85.28 & 19.48 & 44.18 & 45.23 & 84.80 & 85.46 &  19.56 & 46.72 & 47.38\\
\hline
\multirow{2}{*}{\textbf{Ours}} & \multirow{2}{*}{Adaptive RAG}  & \textbf{\M -$cppl$}  & \underline{84.90} & \textbf{89.61} & \underline{20.86} & \underline{53.04} & \underline{53.29} & \underline{87.33} & \underline{92.80}  & \underline{24.65} & \underline{56.94} & \textbf{57.46} \\
 & & \textbf{\M -$uct$} & \textbf{85.63} & \underline{89.46} & \textbf{21.03} & \textbf{53.24} & \textbf{54.98} &  \textbf{87.95} & \textbf{93.15} & \textbf{25.89} & \textbf{57.29} & \underline{56.59} \\
\hline
\multirow{3}{*}{\textbf{Ablations}} & \multirow{3}{*}{Adaptive-RAG}
& {\texttt{w/o Backtrack}} & 83.44 & 88.61 & 20.04 & 51.38 & 52.40  & 86.24  & 90.67 & 22.06 & 52.92 & 53.88 \\
& & {\texttt{w/o Summary}} & 84.82 & 89.00 & 20.72 & 52.11 & 53.07 & 84.79 & 88.47 & 24.48 & 54.39 & 54.72 \\
& & {\texttt{w/o State Monitor}} & 83.75 & 88.79 & 19.82 & 49.44 & 51.07 &  85.27 & 89.04 & 21.40 & 48.42  & 51.86 \\
\hline
\end{tabular}
}
\end{table*}

\subsection{Main Result Analysis (RQ1)}
\label{Main Results}
To answer RQ1, we conduct experiments and report results of the accuracy on the MMCU-Medical, CMB-Exam and CMB-Clin datasets with two LLMs in Table\ref{tab:comparison}. From the reported accuracy, we can find the following observations:

\textbf{Comparison of RAG methods and Base LLMs.} 
Given the complexity of patient queries, we observe that the performance of the Naive RAG methods shows little to no improvement compared to the No RAG baselines. In contrast, the effectiveness of Adaptive RAG is significantly higher than that of Advanced RAG and substantially outperforms other approaches. This underscores the necessity of employing more sophisticated submodules for advanced and adaptive retrieval in domain-specific scenarios.

\textbf{Comparison of \M~and other RAG methods.} 
Our model, \M, clearly outperforms the baselines across all datasets, with an average performance gain of all metrics of up to \textbf{7.20\%}. For instance, the EM and BLEU-4 scores improve by approximately 2.60\%-5.62\% and 8.23\%-13.72\%, respectively. These results highlight the effectiveness of our modules in system state \& memory management, as well as adaptive retrieval.
Additionally, it is worth noting that domain-specific LLMs significantly outperform general LLMs, further demonstrating the importance of pre-training a medical LLM to support \M, in line with \textbf{\textit{C2}}.

\textbf{Strong Generalization Ability.} Table~\ref{tab:general} summarizes the experimental results. \M~demonstrated superior performance compared to the baseline methods in the general domain.
These findings reinforce \M's strong \textbf{generalization ability across diverse domains, tasks, languages, model sizes, and model types}, making it a robust framework for RAG tasks.
 
\subsection{Ablation Study (RQ2)}
\label{Ablation Study}
We perform ablation studies to evaluate the impact of each component within \M, as detailed in Table~\ref{tab:comparison}, with three variants: 
    (1) \M~without \texttt{Backtrack} action (denoted as \texttt{w/o Backtrack}),
    (2) \M~without \texttt{Summary} action (denoted as \texttt{w/o Summary}),
    (3) \M~without State Monitor, relying solely on the LLM's 'Final Answer' action to determine termination, transforming into a black-box system (denoted as \texttt{w/o State Monitor}).

The results reveal that each component contributes positively to the overall performance of \M. The exclusion of any component leads to a noticeable reduction in effectiveness. Particularly, the absence of the State Monitor results in significant performance degradation, highlighting the critical importance of the system state variable in monitoring the process, in line with \textit{\textbf{C1}}, which is essential for preventing overconfidence and ensuring appropriate termination, thereby avoiding excessive or inadequate retrieval.
Moreover, the removal of the \texttt{Backtrack} and \texttt{Summary} actions underscores the necessity of effective memory management. These actions are crucial for mitigating irrelevant noise and maintaining an optimal system state, aligning with the challenges outlined in \textit{\textbf{C3}}.

\subsection{Effiency Study (RQ2)}
\label{EffiencyStudy} 

\begin{table}[t]
  \centering
   \setlength\tabcolsep{2pt}
  \caption{Analysis Comparison of RAG methods. The average duration is computed based on Qwen-32B for the CMB-Exam Dataset.}
  \label{table1}
  \resizebox{1\linewidth}{!}{
  \begin{tabular}{c|c|c|c|c}
    \hline
    \rowcolor[gray]{0.9}
    \textbf{Method} & \textbf{Avg. Interactions} & \textbf{Avg. Retrievers} & \textbf{Avg. Time (s)} & \textbf{Avg. Token} \\
    \hline
    Base & $1$ & $0$ & $7.29$ & $128.89$ \\
    CoT & $1$ & $1$ & $14.37$ & $196.68$ \\
    HyKGE & $2$ & $1$ & $19.76$ & $120.37$ \\
    CoK & $4.31$ & $4.31$ & $125.84$ & $753.02$ \\
    ReACT & $6.84$ & $5.72$ & $130.87$ & $874.29$ \\
    Self-RAG & $5.84$ & $4.22$ & $61.09$ & $567.98$ \\
    FLARE & $4.96$ & $4.96$ & $126.74$ & $785.46$ \\
    DRAGIN & $6.12$ & $6.12$ & $167.95$ & $998.64$ \\
    \cdashline{0-4}
    \textbf{\M} & $4.78$ & $3.37$ & $50.91$ & $458.82$ \\
    \hline
  \end{tabular}
  }
\end{table}

To illustrate the effectiveness of our \M, we conducted a comparative analysis of the time and token overhead between \M~and other RAG approaches based on the Qwen-32B and CMB-Exam datasets, as presented in Table~\ref{table1}.

Our findings reveal that except for models like the Base, which do not require RAG, and those requiring only 1-2 interactions such as CoT and HyKGE, most iterative RAG methods exhibit long average interaction times. Notably, CoK, which lacks adaptive capabilities, is particularly hindered by unrestricted interactions and retrievals, negatively impacting their efficiency. In contrast, methods like ReACT, Self-RAG, FLARE, DRAGIN, and \M~show significantly shorter overheads. For \M, the combination of \textbf{Retrieval and Interaction in the Decoder Stage} and the \textbf{Token Speculative Sampling} strategy results in much lower overhead compared to other adaptive RAG models. This reduced overhead is particularly important, as excessive waiting times can prove to be considerably restrictive, especially in real-world medical Q\&A scenarios where time is a critical factor. 
Therefore, achieving a balance between time overhead and model accuracy is crucial. In this regard, \M~emerges as the most efficient and high-performing framework. In summary, since RAG involves a process of continuous trial and error~\cite{barnett2024seven}, we experimented with numerous strategies and ultimately arrived at \M.

%% file: 6_Conclusion.tex
\section{Conclusions and Future Works}
We introduce the first Turing-Complete RAG system for medical LLMs, named \M. By incorporating a monitored state variable, we have developed a stack memory framework that enables more dynamic and adaptive retrieval processes, effectively addressing the endless and inaccurate retrieval challenges. The \M~framework, with its memory stack system for backtracking and summarizing, effectively reduces the accumulation of erroneous knowledge and irrelevant noise. Our experiments suggest that \M~outperforms existing baselines across multiple real-world medical datasets, showing potential improvements in accuracy and reliability. Furthermore, the successful deployment of \M~on an online platform also highlights its practical value in real-world applications. 
Future efforts will focus on incorporating more complex composed actions, enabling multi-LLM interactions, and exploring its application in other specialized domains.

%% file: 8_Limitations.tex
\section*{Limitations}
Despite the promising results, our framework does have some limitations that need to be addressed. 
First, the computational overhead introduced by the memory stack system and dynamic retrieval processes can be relatively high. While these components are essential for managing complex queries and minimizing error accumulation, they do contribute to increased computational costs. 
In large-scale or real-time applications, this could present challenges related to processing speed and efficiency. Although we have proposed solutions like speculative sampling to alleviate some of this overhead, further optimization is still needed to enhance performance and scalability in time-sensitive environments. 
Another limitation is that the current framework is primarily optimized for general medical applications. While it performs well across a variety of tasks, there remains significant potential for improvement in more specialized medical domains. 
Future work will focus on integrating more complex composed actions, enabling multi-LLM interactions, and expanding the framework's capabilities to accommodate a broader range of specialized medical subfields. This will further enhance its adaptability and performance in these targeted areas.

\section*{Acknowledgments}
This work is supported by the National Natural Science Foundation of China (No.U23A20468).

%% file: 7_Appendix.tex
\newpage
\section{Appendix}


\subsection{The Reasons of Choosing Stack Memory}
\label{reasons for stack memory}
The stack design in \M~has the following five key advantages:
\begin{itemize}[leftmargin=*]
    \item \textbf{Ensuring the Purity of Historical Information}: The stack uses the \texttt{pop} operation to eliminate incorrect or irrelevant knowledge, ensuring that each reasoning step is based on accurate information, preventing the accumulation of erroneous data. This way, the stack remains pure, guaranteeing the reliability and consistency of the reasoning process.
    
    \item \textbf{Efficient Access}: The stack provides $O(1)$ access efficiency, meaning that accessing historical information is done in constant time. In complex query processing, the stack can quickly provide the most recent context or reasoning step, ensuring that the reasoning process is not delayed by information retrieval.
    
    \item \textbf{Storing Multiple Types of Knowledge}: The stack can store both the model's parameterized knowledge via \texttt{Thought} or \texttt{Conclusion} and external non-parameterized knowledge via \texttt{Tool\_Observation} or \texttt{Summary}~\cite{farahani2024decipheringinterplayparametricnonparametric}. This design makes the stack a flexible, multi-level storage structure that integrates multiple sources of knowledge, enhancing the model's reasoning capabilities.
    
    \item \textbf{Flexible Reasoning Control}: The stack dynamically manages the reasoning process. The model decides whether to continue reasoning or conclude based on the state of the information in the stack. By controlling the stack, the model can backtrack, summarize, or terminate reasoning when necessary, improving the flexibility and controllability of the reasoning process and avoiding over-reasoning or premature conclusions.
    
    \item \textbf{Human-like Reasoning}: The stack's "Last In, First Out" (LIFO) nature effectively simulates the way human experts (i.e. doctors) reason during medical diagnoses or decision-making. Experts typically accumulate information step by step and adjust their conclusions when new evidence arises. This design of the stack makes the model's reasoning process more aligned with real-world decision-making, enabling it to handle multi-step reasoning and complex decisions more effectively.
\end{itemize}

\subsection{Full Proof of Turing-Complete}
\label{Turing-Complete}
We prove this by showing that for any Turing machine $T$, there exists our \textsc{Tc} that can simulate $T$:

Let $\mathbb{N}$ denote the set of natural numbers, and $\mathbb{R}$ the set of real numbers. We consider computations over finite alphabets and prove the Turing completeness of our proposed model.

\paragraph{\textbf{Definition 1. (Standard Turing Machine).}}
A standard Turing Machine is a 7-tuple Turing machine $T = (\mathcal{S}_T, \mathcal{A}_T, \mathcal{M}_T, \delta_T, st_0, st_{accept}, st_{reject})$ where:
\begin{itemize}
    \item $\mathcal{S}_T$ is a finite set of states.
    \item $\mathcal{A}_T \subset \mathcal{M}_T$ is the input alphabet.
    \item $\mathcal{M}_T$ is the tape alphabet, with $\sqcup \in \mathcal{M}_T \setminus \mathcal{A}_T$ as the blank symbol.
    \item $\delta_T: \mathcal{S}_T \times \mathcal{M}_T \to \mathcal{S}_T \times \mathcal{M}_T \times \{L, R\}$ is the transition function.
    \item $st_0 \in \mathcal{S}_T$ is the initial state, and its initial value and upper bound is \$Large\_Value\$.
    \item $st_{accept}, st_{reject} \in \mathcal{S}_T$ are the accept and reject states, respectively.
\end{itemize}

\paragraph{\textbf{Definition 2. ($T$ Configuration).}}
A configuration of $T$ is a tuple $c_{T} = (st, w_1aw_2)$, where:
\begin{itemize}[leftmargin=*]
    \item $st \in \mathcal{S}_T$ is the current system state value,
    \item $w1, w2$ represents the sequence of actions $a\in \mathcal{A}_T$ to the left / right of the tape head.
\end{itemize}

\paragraph{\textbf{Definition 3. (\textsc{Tc} Configuration).}}
A configuration of \textsc{Tc} is a tuple $c_{\textsc{Tc}} = (s, w, \mathcal{M}, \{halt~ ||~ continue\}) \in \mathcal{S} \times \mathcal{A} \times \mathcal{M} \times \{halt, continue\}$, where:
\begin{itemize}[leftmargin=*]
    \item $s \in \mathcal{S}$ is the current system state value, and its initial value and upper bound is \$Large\_Value\$.
    \item $w$ is the remaining input actions,
    \item $\mathcal{M}$ is the current stack content,
    \item $\{halt ~||~ continue\}$ is the conditions that determine whether the system will continue or halt.
\end{itemize}

\paragraph{\textbf{Definition 4. (Computation Step).}}
For configurations of \textsc{Tc} at step $t$ $c_{t} = (s_t, w_t, \mathcal{M}_t, s_t\ge \sigma)$ and $c_{t+1} = (s_{t+1}, w_{t+1}, \mathcal{M}_{t+1}, \cdot)$, we say $c_t \vdash_{\text{\textsc{Tc}}} c_{t+1}$ if and only if:
\begin{enumerate}
    \item $w_t = a w_{t+1}$ for some $a \in \mathcal{A} \cup \{\varepsilon\}$
    \item $\delta(s_t, a) = (s_2, op, b)$
    \item $\mathcal{M}_{t+1} = \begin{cases}
        push(\mathcal{M}_t, b) & \text{if } op = push \\
        pop(\mathcal{M}_t) & \text{if } op = pop
    \end{cases}$
    \item The computation terminates if $s_{t+1} < \sigma$
\end{enumerate}
where $top(\mathcal{M})$ returns the top element of the stack $\mathcal{M}$.

Next, we will break down this proof process into the following steps:
\paragraph{Step 1: Construction of the Turing System. }
Given a Turing machine $T = (\mathcal{S}_T, \mathcal{A}_T, \mathcal{M}_T, \delta_T, st_0, st_{accept}, st_{reject})$, we interpret $\text{\textsc{Tc}} = (\mathcal{S}, \mathcal{A}, \mathcal{M}, \delta, s_0, \mathcal{F}, \sigma)$ with the following components:

\begin{enumerate}[leftmargin=*,noitemsep,topsep=2pt]
\item $\mathcal{S} = \mathcal{S}_T \cup {f}$, where $f \notin \mathcal{S}_T$ is the end state.
\item $\mathcal{A} = \mathcal{A}_T$, $\mathcal{M} = \mathcal{M}_T \cup \mathcal{A}$, $s_0 = st_0$, $\mathcal{F} = \{f\}$, $\sigma$ is the hyper-threshold to decide accept or reject.
\item The transition function $\delta$ is defined as:
\end{enumerate}
\begin{footnotesize}
\begin{align*}
\delta(s,a)=\begin{cases}
        (s',push,s\geq\sigma)&\text{if } \delta_T(st, a)=(st',a,R) \\
        (s',pop,s\geq\sigma)&\text{if } \delta_T(st,a)=(st',a,L) \\
        (f,no\_op,a, s<\sigma)&\text{if } st\in\{st_{accept},st_{reject}\}
    \end{cases}
\end{align*}
\end{footnotesize}
where $a$ is the executing action for $T$ or \textsc{Tc}. The operations $push$ and $pop$ are corresponding to the right ($R$) and left ($
L$) movements in $T$. Transition function $\delta(\cdot,\cdot)$ is for \textsc{Tc} and $\delta_T(\cdot,\cdot)$ is for $T$. The operation $no\_op$ indicates no operation is performed when the system is halting. This $\delta(\cdot, \cdot)$ decides the next state value and action ($R$, $L$, $no\_op$).


\paragraph{Step 2: Configuration Mapping}
Next, we define a bijective mapping function $h$ that maps each configuration of $T$ to \textsc{Tc}. 
We now define the configuration $c$ of $T$ as $c=(st, w_1 a w_2)$, $w_1, w_2$ represents the sequence of actions to the left / right of the tape head, where $st \in \mathcal{S}_T$, $a \in \mathcal{A}_T$ is the tape head/action.
The configuration of \textsc{Tc} $c_{\textsc{Tc}
}$ is composed of the tuple as (system state, remaining actions, stack memory actions, whether to halt) as:
\[ h(st, w_1aw_2) = (s, w_2, w_1^Ra, s\geq \sigma) = c_{\textsc{Tc}
}\]
where $w_1^R$ is the reverse of $w_1$, which is necessary due to Last-In-First-Out (LIFO) principle of $\mathcal{M}$.

\paragraph{Step 3: Simulation of Computation Steps}
We now prove that \textsc{Tc} can simulate each step of $T$ as in Lemma~\ref{lemma 1}:
\begin{lemma}
\label{lemma 1}
If $c_1\vdash_T c_2$ in $T$, then $h(c_1)\vdash_{\textsc{Tc}}^* h(c_2)$ in \textsc{Tc}, where $*$ denotes one or multiple steps of derivation, $\vdash$ is the action shift operator.
\end{lemma}

\begin{proof}
Let $c_1 = (st_1, w_1a_1w_2)$ and $c_2 = (st_2, w_1'a_2w_2')$ be configurations of $T$ with $a_t=\{L,R\}$. We prove Lemma~\ref{lemma 1} by considering the two cases $push$ and $pop$ corresponding to the possible directions of tape head movement in $T$.
\begin{itemize}
[leftmargin=*,noitemsep,topsep=2pt]
    \item \textbf{Case 1}: If $\delta_T(st_1, a_1) = (st_2, a_2, R)$, then:
    \begin{align*}
    \customfootnotesize
    h(c_1) &= (st_1, w_2, w_1^Ra_1, st_1\geq\sigma) \\
    &\vdash_{\textsc{Tc}} (s_2, {w_2'}, w_1^Ra_1a_2, s_2\geq\sigma) \quad (\text{\textbf{push} $a_2$})\\
    &= h(st_2, \underline{w_1a_1}{a_2w_2'})= h(st_2, \underline{w_1'}a_2w_2')= 
    h(c_2)
    \end{align*}
    
    \item \textbf{Case 2}: If $\delta_T(st_1, a_1) = (st_2, a_2, L)$, then:
\end{itemize}
\begin{footnotesize}
    \begin{align*}
    \customfootnotesize
    h(c_1) &= (st_1, w_2, w_1^Ra_1, st_1\geq\sigma) \\
    &\vdash_{\textsc{Tc}} (s_2, w_2, w_1^R, s_2\geq\sigma) \quad (\text{\textbf{pop} $a_1$}) \\
    &\vdash_{\textsc{Tc}} (s_2, w_2', w_1'^R a_2, s_2\geq\sigma) \quad \text{(\textbf{push} $a_2$, if $a_2 \neq \text{null}$)} \\
    &= h(st_2, w_1'a_2w_2') = h(c_2)
    \end{align*}
\end{footnotesize}
\end{proof}
It is noted that the two cases correspond to the meta operations of \texttt{push} (Case 1) and \texttt{pop} (Case 2) in \M.

\paragraph{Step 4: Preservation of Acceptance and Rejection}

\begin{lemma}
$T$ accepts (rejects) the whole input $w$ if and only if \textsc{Tc} reaches a configuration $(f, NULL, \mathcal{M}, f<\sigma)$ from the initial configuration $(s_0, w, NULL, s_0\geq \sigma)$.
\end{lemma}

\begin{proof}
We prove the consistency of reaching the termination state from both $T\to \textsc{Tc}$ and $\textsc{Tc} \to T$ perspectives.

(\textbf{Forward} $\Rightarrow$) Assume $T$ accepts (rejects) input $w$. Then:
$\exists t \in \mathbb{N}$ such that after $t$ steps, $T$ enters state $st_{accept}$ (or $st_{reject}$). Then, Let $(st_t, w_t)$ be the configuration of $T$ at step $t$, where $st_t \in \{st_{accept}, st_{reject}\}$. By our construction of $\delta$, $\forall a \in \mathcal{A}$, for $\forall st \in \{st_{accept}, st_{reject}\}$ we have:
    \[\delta(s, a) = (f, no\_op, a, f<\sigma)\]
Therefore, if $h(st_t, w_t) = (s_t, w_2, w_1^Ra, s_t \geq \sigma)$ is the corresponding configuration in \textsc{Tc}, then:
    \[(s_t, w_2, w_1^Ra, s_t\geq \sigma) \vdash_{\textsc{Tc}}^t (f, NULL, \mathcal{M}, f<\sigma)\]
Thus, \textsc{Tc} achieve $f$ with system state value lower than $\sigma$.

(\textbf{Backward} $\Leftarrow$) Assume \textsc{Tc} reaches configuration $(f, NULL, \mathcal{M}, f<\sigma)$ from $(s_0, w, NULL, s_0\geq \sigma)$. Then:
$\exists t \in \mathbb{N}$ such that:
\[(s_0, w, NULL, s_0\geq \sigma) \vdash_{\textsc{Tc}}^t (f, NULL, \mathcal{M}, f<\sigma).\] 
Let $(s_{t-1}, w_2, w_1^Ra, s_{t-1} \geq \sigma)$ be the configuration of \textsc{Tc} at step $t-1$. Then, by the construction of $\delta$, the only way to reach $f$ is if $s_{t-1} \in \{st_{accept}, st_{reject}\}$.
Therefore, the corresponding configuration of $T$ at step $t-1$ must be $(st_{t-1}, w_1aw_2)$, where $st_{t-1} \in \{st_{accept}, st_{reject}\}$.
Thus, $T$ must have entered $st_{accept}$ or $st_{reject}$, implying that $T$ accepts or rejects $w$.

From both the \textbf{Forward} and \textbf{Backward} proof, $T$ accepts (rejects) input $w$ if and only if \textsc{Tc} reaches a configuration $(f, NULL, \mathcal{M}, f<\sigma)$ from the initial configuration $(s_0, w, NULL, s_0\geq \sigma)$.
\end{proof}

By the above lemmas, we have shown that: (1) \textsc{Tc} can simulate every move of $T$. (2) \textsc{Tc} halts if and only if $T$ achieve acceptance or rejection behavior. \textbf{Thus, \textsc{Tc} can simulate any computation of any Turing machine $T$, which by definition makes \textsc{Tc} Turing complete}, which is consistent with the LLM's backbone——transformers are Turing-Complete~\cite{sss3}.

\subsection{Convergence Analysis of State Value}
\label{appendix:convergence}

In this section, we rigorously analyze the convergence behavior of the \M~model. We focus on two key aspects: the role of information gain in reducing uncertainty and the application of Lyapunov stability theory to establish finite-time convergence.
To rigorously prove the finite-time convergence of the \M~process, we analyze its dynamics using Lyapunov stability theory.

\paragraph{Definition of the Lyapunov Function}

We define a Lyapunov function that quantifies the deviation of the system state \( st_t \) from a convergence threshold \( \sigma \):
\begin{equation}
    V(st_j) = \max(0, st_j - \sigma),
\end{equation}
where \( \sigma \) is a predefined threshold indicating sufficient certainty about the final answer \( Y \). When \( st_j \leq \sigma \), the system is considered to have converged.

\paragraph{Properties of the Lyapunov Function}

The function \( V(st_j) \) possesses the following properties:
\begin{enumerate}
    \item \textbf{Non-negativity:} By definition, \( V(st_j) \geq 0 \) for all \( j \geq 0 \), and \( V(st_j) = 0 \) if and only if \( st_j \leq \sigma \).
    \item \textbf{Monotonic Decrease:} Our objective is to prove that \( V(st_j) \) decreases over time, i.e.,
    \begin{equation}
        V(st_{j+1}) - V(st_j) \leq 0.
    \end{equation}
\end{enumerate}
Next, first we will provide the \textit{Action Validity Assumption}, and then we will prove these two properties respectively.

\subsubsection{Assumption of Action Validity}

\textbf{\textit{Assumption 1 (Action Validity):}} \textit{Leveraging the robust task comprehension capabilities of LLMs~\cite{ton2025understanding}, we assume that every correctly executed action—such as \texttt{push} and \texttt{pop} operations—yields meaningful information gain. This gain enhances the system's ability to accurately predict the final answer. In contrast, when an incorrect action occurs, the system employs a \texttt{pop} operation to discard the erroneous information. This mechanism ensures that only relevant and accurate content is retained throughout the retrieval-augmented and reasoning processes.}

\subsubsection{Uncertainty Reduction}

To formalize the reduction in uncertainty over time, we analyze the system's behavior through the lens of mutual information. Let \( Y \) denote the final answer and \( st_j \) represent the memory state at step \( j \). The mutual information between \( Y \) and \( st_j \) is defined as:
\begin{equation}
    I(Y; st_j) = \mathbb{E}_{p(y, st_j)} \left[ \log \frac{p(y, st_j)}{p(y)p(st_j)} \right].
\end{equation}

Our goal is to show that after each action \( a_t \), the uncertainty regarding \( Y \) is reduced. More precisely, we aim to demonstrate:
\begin{align}
    \mathbb{E} \left[ \log p(Y \mid st_j) \right] - \mathbb{E} \left[ \log p(Y \mid st_{j-1}) \right] \\ = I(Y; st_j \mid st_{j-1}).
\end{align}
The expectation \(\mathbb{E} \left[ \log p(Y \mid st_j) \right]\) measures how well the model fits the true labels given the state \(st_j\), representing the expected log-likelihood. This value reflects the information gain of the content at step \(j\) with respect to \(Y\).  
If we can show that this value at step \(j\) is higher than at step \(j-1\), it would indicate that the model's inference is progressively approaching the true distribution. 
Thus, we assume that the state transitions follow a Markov process, where $st_j$ depends only on $st_{j-1}$ and the previous state could be observed $p(st_{j-1}|st_j)=1$. This Markov structure also implies that the observation \(Y\) is conditionally independent of the previous state \(st_{j-1}\) given the current state \(st_j\), expressed formally as \(Y \,\perp\, st_{j-1} \mid st_j\). This leads to the following:

\paragraph{Proof:}
\begin{align}
    &\mathbb{E} \left[ \log p(Y \mid st_j) \right] - \mathbb{E} \left[ \log p(Y \mid st_{j-1}) \right] \\
    &= \mathbb{E} \left[ \log \frac{p(Y \mid st_j)}{p(Y \mid st_{j-1})} \right] \nonumber \\
    &= \mathbb{E} \left[ \log \frac{p(Y, st_j \mid st_{j-1})}{p(Y \mid st_{j-1}) \, p(st_j \mid st_{j-1})} \right] \nonumber \\
    &= I(Y; st_j \mid st_{j-1}).
\end{align}

In summary, the behavior of the conditional mutual information after each action \( a_t \) is as follows:
\begin{equation}
    I(Y; st_j \mid st_{j-1}) = 
    \begin{cases} 
      > 0, & \text{if } a_t = \texttt{push}, \\[6pt]
      \geq 0, & \text{if } a_t = \texttt{pop}.
    \end{cases}
\end{equation}

Thus, with every action, mutual information either increases or remains non-decreasing, ensuring a progressive reduction in uncertainty. Specifically, \texttt{push} operations incorporate task-relevant knowledge, while \texttt{pop} operations remove noisy or irrelevant information.

\subsubsection{Lyapunov Stability and Finite-Time Convergence of \M}

\paragraph{Impact of Actions on the Lyapunov Function}

Each action \( a_t \) (whether \texttt{push} or \texttt{pop}) influences the system’s uncertainty, and consequently, the Lyapunov function, which implies that the Lyapunov function satisfies:
\begin{align}
    & V(st_{j+1}) - V(st_j) \\ & = \max(0, st_{j+1} - \sigma) - \max(0, st_j - \sigma) \\& \leq st_{j+1} - st_j \leq 0.
\end{align}

Thus, the Lyapunov function is non-increasing, ensuring that the system state progressively approaches the convergence threshold.

\paragraph{Proof of Finite-Time Convergence}

To establish finite-time convergence, we assume that the decrease in \( V(st_j) \) after each action is bounded below by a positive constant. That is, there exists an \( \epsilon > 0 \) such that for all \( t \):
\begin{equation}
    V(st_{j+1}) - V(st_t) \leq -\epsilon.
\end{equation}

\textbf{Theorem 1 (Finite-Time Convergence).} There exists a finite time \( T \) such that:
$
    V(st_T) = 0.
$

\textbf{Proof.} (By contradiction) Suppose that \( V(st_j) > 0 \) for all \( t \). Then, due to the bounded decrease, for any \( n \in \mathbb{N} \),
\begin{equation}
    V(st_n) \leq V(st_0) - n\epsilon.
\end{equation}
Choosing \( n > \frac{V(st_0)}{\epsilon} \) yields:
$
    V(st_n) < 0,
$
which contradicts the non-negativity of \( V(st_j) \). Hence, there must exist a finite \( T \) such that \( V(st_j) = 0 \).

\paragraph{Convergence Time Bound}
The convergence time \( T \) satisfies:
\begin{equation}
    T \leq \left\lceil \frac{V(st_0)}{\epsilon} \right\rceil.
\end{equation}


Alternatively, if the decrease is proportional, i.e.,
$
V(s_{t_{j+1}}) \leq (1 - \beta)V(s_{t_j}),
$
for some \( \beta > 0 \), then by repeated application we obtain
$
V(s_{t_j}) \leq V(s_{t_0})(1 - \beta)^T.
$
To guarantee that \( V(s_{t_j}) \leq \delta \) for a small threshold \( \delta \), it is necessary that
\[
T \geq \frac{\log(\delta / V(s_{t_0}))}{\log(1 - \beta)}.
\]

However, experimental observations indicate that the behavior of \( V(s_t) \) actually unfolds in two stages. Initially, the decay is exponential in nature, which can be modeled as
$
V(s_t) = V(s_{t_0}) e^{-\lambda t},
$
so that ensuring \( V(s_t) \leq \delta \) requires
\[
T \geq \frac{\ln(V(s_{t_0})/\delta)}{\lambda}.
\]
After this initial phase, the decay tends to become nearly linear. Therefore, the overall time cost \( T \) is estimated to be composed of these two parts: an initial exponential decay phase followed by a subsequent nearly linear decay phase.

\paragraph{Conclusion}

By defining the Lyapunov function \( V(st_j) = \max(0, st_j - \sigma) \) and carefully analyzing the effects of both \texttt{push} and \texttt{pop} operations, we have demonstrated that the uncertainty in the \M~process decreases over time. Moreover, under the assumption of a bounded decrease per step, the system is guaranteed to converge to the target state within a finite number of steps. This theoretical framework underscores the robustness of the \M~and highlights the importance of effectively managing information gain to achieve both efficiency and accuracy.

\subsection{Review of LLMs}
\label{Review of LLMs}
\paragraph{\textbf{Definition (Large Language Models)}.}
Generative LLMs are powerful language models capable of generating coherent and contextually relevant text. Through pretraining on large-scale text corpora and alignment fine-tuning to follow human instructions, they can generate human-like text based on given prompts or inputs.
Typically, LLMs $\Theta$ model the probability of a sentence (\textit{i.e.}, a sequence of word tokens) $l=\left(w_1, w_2, \ldots, w_n\right)$ as $P(s  ;\Theta)=\prod_{i}^n P\left(w_i \mid w_{<i};\Theta \right)$, where $w_i$ denotes the $i$-th token $w_i$ of the sentence $l$ and $w_{<i}$ denotes the partial word token sequence before the $i$-th step.

\subsection{Model Pretrain}
\label{sec:model pretraining}

\subsubsection{Pretrain LLMs}
The high-quality pre-training corpus can greatly improve the performance of LLM and even break the scaling laws to some extent \cite{gunasekar2023textbooks,ding20243ds}. Among them, continuous pre-training is a crucial phase where the language model undergoes extensive training on vast and diverse unlabeled datasets. This process spans multiple iterations, each aimed at refining the model's language understanding capabilities~\cite{luo2024kuaijichineseaccountinglarge}. 
Initially, the LLM is initialized with the pre-trained weights of basic LLMs and learns to predict missing words or segments within sentences using self-supervised learning objectives such as masked language modeling (MLM) and next-sentence prediction (NSP). Through exposure to a wide variety of textual data sources, the model gradually acquires a rich domain understanding of language structure, semantics, and context in the medical domain. Additionally, techniques like attention mechanisms and multi-layer architectures are employed to capture complex linguistic patterns and dependencies. 

It should be noted that we did not use supervised fine-tuning because it would lead to overconfidence in the large model, resulting in the large model often receiving diagnostic results directly without planning and calling professional medical knowledge retrieval tools.
\begin{table}[h]
\centering 
\caption{Medical pre-train data statistics}
\label{tab:pretraindata}
\begin{tabular}{@{}cccc@{}}

\toprule[0.75 pt]
Type                   & Dataset                & Size          \\ \midrule[0.75pt]
\multirow{2}{*}{Dialogues}   & RealHospital-QA                & 2318 MB           \\
                        & Web-QA             & 567 MB                    \\ \midrule[0.75pt]
\multirow{1}{*}{Knowledge graphs} & Medical-KG              & 379 MB           \\

                        \midrule[0.75pt]
\multirow{1}{*}{Exams} & Medical-Exam              & 443 MB           \\

                        \midrule[0.75pt]

\multirow{2}{*}{Textbooks} &  Chinese-Textbook             & 52 MB  \\
                            & English-Textbook         & 212 MB        
                           \\ \midrule[0.75pt] 
Guidelines                  & Med-Guidelines         & 878 MB    
                        \\ \midrule[0.75pt]
Encyclopedia                  & Med-Encyclopedia         & 798 MB    
                        \\ \midrule[0.75pt]
\multicolumn{2}{c}{Total}                            & \textbf{5647 MB} \\ \bottomrule[0.75 pt]
\end{tabular}
\end{table}
\subsubsection{Data Preparement}
To build a diverse medical corpus, we compiled data from multiple sources, ensuring broad coverage across various medical domains.

\begin{itemize}
    \item \textbf{Dialogues}: The \textbf{RealHospital-QA} dataset includes real-world clinical conversations, while \textbf{Web-QA} provides Q\&A pairs from online health forums, capturing common public inquiries.
    \item \textbf{Knowledge Graphs}: \textbf{Medical-KG} organizes medical knowledge into entities and relationships, integrating data from clinical guidelines, research papers, and textbooks.
    \item \textbf{Exams}: The \textbf{Medical-Exam} dataset consists of questions from medical exams, aiding the model in handling complex diagnostic scenarios.
    \item \textbf{Textbooks}: We included \textbf{Chinese-Textbook} and \textbf{English-Textbook} datasets to provide foundational knowledge in both languages.
    \item \textbf{Guidelines}: \textbf{Med-Guidelines} comprises official medical guidelines, essential for evidence-based practice.
    \item \textbf{Encyclopedia}: The \textbf{Med-Encyclopedia} offers concise explanations of medical terms and conditions.
\end{itemize}

These datasets span multiple specialties, giving the model a comprehensive understanding of medical knowledge. The total corpus size is 5647 MB, as shown in Table \ref{tab:pretraindata}.
\subsubsection{Pretrain Loss Function}
The pertaining loss function is defined as follows:
\begin{equation}
\mathcal{L}_{\textbf{pretrain}} = \sum_{t \in \text{masked}} \log P(w_t \mid l_{\backslash t}; \Theta),
\end{equation}
where $l_{\backslash t}$ represents the remaining part of the sequence $l$ after masking the \(t\)-th word. 
\subsubsection{Training Setup}
We performed the continuous pre-training on Qwen1.5-32B-Chat using the aforementioned medical datasets. The pre-training was conducted on eight H100 GPUs for one epoch with a learning rate of 1e-4, and the entire training process spanned 4 days. This configuration was chosen to balance computational efficiency with the need for thorough learning, allowing the model to effectively internalize the extensive medical knowledge embedded in the training data. Through this process, we aimed to enhance the model's reasoning and planning capabilities, ensuring it can provide accurate and reliable medical analysis in real-world applications.


\subsection{Prompt and Algorithm}
\label{sec: prompts and algorithm} 
\subsubsection{Prompt Format.} In this module, we will provide a detailed introduction to the Prompt used in our entire model in the following Prompt:

\subsubsection{\M~Algotithm.} Algorithm~\ref{alg:reasoning_loop} describes the reasoning loop of \M~for generating a final answer based on user query using a pre-trained Medical LLM. The process begins by initializing the stack memory and the initial state (Lines 1-2). The user query is then pushed into the stack memory (Line 3).

Within the loop, which continues until the action limit is reached or the system state value drops below the threshold $\sigma$, the model generates an action based on the current memory stack (Lines 5-6). If the action type is identified as a \texttt{Conclusion}, the system state is recalculated; if it remains within acceptable bounds, the final answer is confirmed and pushed into the stack, terminating the loop (Lines 7-11). Otherwise, the final answer is reclassified as a \texttt{Thought}, and processing continues (Lines 12-14).

If the action type is a \texttt{Thought}, the system state is updated, and the thought result is pushed into the stack (Lines 15-17). For \texttt{Tool\_Use}, the model retrieves relevant observations using specified tools or a knowledge base, which are then pushed into the stack (Lines 18-20). When the action type is \texttt{Backtrack}, the top of the stack is popped, and the previous system state is restored (Lines 21-23). If the action is a \texttt{Summary}, the stack is adjusted by popping irrelevant content, summarizing it, and pushing the summary back onto the stack (Lines 24-26).

Finally, after exiting the loop, the top of the stack is returned as the final answer (Lines 27-28).

\begin{algorithm}[htbp]
\footnotesize
    \caption{\M~Inference Algorithm}
    \label{alg:reasoning_loop}
    \begin{algorithmic}[1]
    \Require ${User\_Query}$, pretrained Medical LLM $\Theta$, knowledge base $\mathcal{D}$, system value threshold $\sigma$, \M~Prompt $\mathcal{P}$
    \Ensure ${Conclusion}$
    \State Initialize stack memory $\mathcal{M}$, state $s\leftarrow \text{\$Large\_Value\$}$
    \State \texttt{Push} ${User\_Query}$ into $\mathcal{M}$
    \State $actions\_taken \leftarrow 0$
    \While{$actions\_taken < max\_loop$ \textbf{and} $s \ge \sigma$}
        \State ${Action} \leftarrow \Theta(\mathcal{M}~|~\mathcal{P})$

        \If{$Action$ is \texttt{Conclusion}}
            \State $f \leftarrow \text{calculate\_state}({User\_Query}, {Action})$
            \If{$f < \sigma$}
                \State \texttt{Push} $Action$ into $\mathcal{M}$
                \State \textbf{Break}
            \Else
                \State $action\_type \leftarrow \texttt{Thought}$, $s\leftarrow f$
                \State \texttt{Push} $Action$ into $\mathcal{M}$
            \EndIf
        \ElsIf{$Action$ is \texttt{Thought}}
            \State $s \leftarrow \text{calculate\_state}({User\_Query}, {Action})$
            \State If $s < \sigma$ Then $s\leftarrow \sigma$
            
            \State \texttt{Push} $Action$ into $\mathcal{M}$
        \ElsIf{$Action$ is \texttt{Plan}}
            \State \texttt{Push} $Action$ into $\mathcal{M}$
        \ElsIf{$Action$ is \texttt{Tool\_Use}}
            \State $observation \leftarrow \text{call}(tool\_type)$ \& search $\mathcal{D}$
            \State \texttt{Push} $observation$ into $\mathcal{M}$
        \ElsIf{$Action$ is \texttt{Backtrack}}
            \State $old\_action$ = \texttt{Pop} from $\mathcal{M}$
            \State If $old\_action$ is \texttt{Thought} then reset $s$ to the previous state
        \ElsIf{$Action$ is \texttt{Summary}}
            \State \texttt{Pop} from $\mathcal{M}$
            \State \texttt{Push} $Action$ into $\mathcal{M}$
        \EndIf

        \State $actions\_taken \leftarrow actions\_taken + 1$
    \EndWhile
    
    \State \textbf{Return} $Conclusion \leftarrow$ top($\mathcal{M}$)
\end{algorithmic}
\end{algorithm}

\newpage
\begin{tcolorbox}
[colback=lightgray!20,colframe=darkgray!80,title= \M~Prompt]
\label{tab:prompt}
Answer the following questions as best you can. You have access to the following tools:
\newline
\newline
\textbf{\textit{[Insert tool\_descriptions here]}}

\textit{i.e.} DOC\_RAG: You can obtain medical knowledge from authoritative documents via this tool to help you reply.
\newline
\newline
Please think strictly according to the provided way of thinking without omission, and use the following format:
\newline
\newline
\textbf{\textit{[Actions Pipeline Format]}}

\texttt{User\_Query}: User's questions or observed information,

\texttt{Thought}: You should think about what to do, whether to answer questions based on the results of the tool or decide which tool to use.

\texttt{Tool\_Use}: The tool to be used must be one of [{tool\_name}], do not add any extra characters or symbols! Only the name of the tool can be output!

\texttt{Tool\_Observation}: The answer provided by the tool (not generated by you)

\texttt{Plan}: Make decisions to outline subsequent actions or strategies.

\texttt{Summary}: When the previous action outputs a large amount of vocabulary and you need to summarize it, please output a detailed summary based on your knowledge.

\texttt{Backtrack}: When the result of the previous action is meaningless to your task and you need to re-execute it, apply this.

...(\texttt{Thought}/\texttt{Tool\_Use}/\texttt{Summary}/\texttt{Backtrack} here can be repeated zero or more times)

\texttt{Thought}: I now know the final answer.
\texttt{Conclusion}: the final answer to the original input question.
\newline
\newline
\textbf{\textit{[Start Conversation]}}

Begin!

\texttt{User\_Query:}...
\end{tcolorbox}

\subsection{Detailed Experimental Setup}
\label{Detailed Experimental Setup}

\subsubsection{Dataset.}
Our experiments are conducted on two open-source query sets: MMCU-Medical~\cite{mmcu} and CMB-Exam~\cite{cmb} datasets, which are designed for multi-task Q\&A and encompass single and multiple-choice questions in the medical field,
and one open-domain Q\&A dataset CMB-Clin~\cite{cmb} which is the inaugural multi-round question-answering dataset based on real, complex medical diagnosis and treatment records. 
For MMCU-Medical, the questions are from the university medical professional examination, covering the three basic medical sciences, pharmacology, nursing, pathology, clinical medicine, infectious diseases, surgery, anatomy, etc., with a total of 2,819 questions.
The CMB-Exam dataset utilizes qualifying exams as a data source in the four clinical medicine specialties of physicians, nurses, medical technicians, and pharmacists, with a total of 269,359 questions. The CMB-Clin dataset contains 74 high-quality, complex, and real patient cases with 208 medical questions.

\subsubsection{RAG Tools.} We have involved several types of RAG tools:

\textbf{(1) Knowledge Graph Search. }
\textit{CMeKG} (Clinical Medicine Knowledge Graph)\footnote{https://cmekg.pcl.ac.cn/}\footnote{https://github.com/king-yyf/CMeKG\_tools} ~\cite{cmekg}, \textit{CPubMed-KG} (Large-scale Chinese Open Medical Knowledge Graph) \footnote{https://cpubmed.openi.org.cn/graph/wiki} and \textit{Disease-KG} (Chinese disease Knowledge Graph)\footnote{https://github.com/nuolade/disease-kb} are open-source medical KGs, which integrates extensive medical text data, including diseases, medications, symptoms and diagnostic treatment technologies. 
The fused KG has 1,288,721 entities and 3,569,427 relations. However, due to the lack of medical entity descriptions in its entities, we collect relevant entity knowledge from Wikipedia\footnote{https://www.wikipedia.org/}, Baidu Baike\footnote{https://baike.baidu.com/}, and Medical Baike\footnote{https://www.yixue.com/}, and store them as entity descriptions. As mentioned by HyKGE~\cite{HyKGE} and GraphRAG~\cite{GraphRAG}, we choose the reasoning chains and knowledge communities as knowledge carriers.

\textbf{(2) Documents Search. } We have collected over 2 million medical documents of 3B size from drug instructions, medical textbooks, medical encyclopedias, clinical diagnosis and treatment guidelines, medical papers, and medical electronic medical records. To be specific, we utilize the GTE embedding model~\cite{gte} ``gte\_sentence-embedding''\footnote{https://www.modelscope.cn/models/damo/nlp\_gte\_sen\\tence-embedding} to obtain the embedding for each document, which is currently the top-performing model for text vector embedding in the retrieval field. We set the document chunk size to 128 with overlap size 50.

\textbf{(3) Web and Pedia Search. } In order to support the implementation of online retrieval, we also support searching on encyclopedias such as Wikipedia and MedNet (for this purpose, we pre-trained a W2NER model~\cite{w2ner} for medicine to extract medical entities). In addition, we also use search engines such as Bing and Google for web retrieval.

\textbf{(4) Electronic Medical Record Database Search. } In order to support the retrieval of similar patient information, we also apply MIMIC-III~\cite{johnson2016mimic}, MIMIC-IV~\cite{johnson2023mimic}, and eICU~\cite{pollard2018eicu} datasets, following the code translation ``ICD9-CM''~\cite{9992086,xu2024protomix} principle to support the retrieval of similar patients.

\subsubsection{LLM Turbo.}
To fairly verify whether \M~can effectively enhance LLMs, we selected the general-domain large model and the medical-domain large model as the base models and explored the gains brought by \M: Qwen-1.5-32B-chat.
In the general domain, we select Llama3.2-1B-Instruct, Llama3.1-8B-Instruct, Qwen2.5-7B-Instruct and Qwen2.5-14B-Instruct for comparison.å

\subsubsection{Compared Methods.}
In order to explore the advantages of the \M, we compare the \M~results against twelve other models: 
(1) \textbf{Base Model (Base)} servers as the model without any external knowledge, used to check the improvement effect of different RAG methods. We use Qwen and pre-trained ones as base models.
(2) \textbf{CHAIN-OF-THOUGHT (CoT)~\cite{wei2023chainofthought}} generates a series of intermediate reasoning steps to perform complex reasoning.
(3) \textbf{Single Round-RAG (SR-RAG)} is selected with the combination of KGRAG~\cite{soman2023biomedical,kgrag_arxiv,Sen2023KnowledgeGL}, embedding-based Document RAG and web search based
on the initial question.
(4) \textbf{Fix Length RAG (FL-RAG)~\cite{khandelwal2020generalizationmemorizationnearestneighbor,borgeaud2022improvinglanguagemodelsretrieving,ram2023incontextretrievalaugmentedlanguagemodels}} triggers the retrieval module every $n$ tokens and the tokens generated in the previous token window are utilized as the query.
(5) \textbf{Fix Sentence RAG (FS-RAG)~\cite{trivedi2023interleavingretrievalchainofthoughtreasoning}}: Similar to FL-RAG, we retrieves based on every sentence.
(6) \textbf{CHAIN-OF-NOTE (CoK)~\cite{li2023chainofknowledge}} generates sequential thoughts after retrieved knowledge, enabling a thorough evaluation of their relevance to the given question and integrating these thoughts to formulate the final answer.
(7) \textbf{Summarizing Retrievals  (SuRe)~\cite{chen2024improving}} constructs summaries of the retrieved passages for each of the multiple answer candidates and confirms the most plausible answer from the candidate set by evaluating the validity and ranking of the generated summaries. 
(8) \textbf{Hypothesis Knowledge Graph Enhanced Framework (HyKGE)~\cite{HyKGE}} leverages the hypothesis output and knowledge graph to enhance model inference. 
(9) \textbf{Reasoning And Acting (ReACT)~\cite{react}} overcomes issues of hallucination and error propagation prevalent in chain-of-thought reasoning ahd actions.
(10) \textbf{Self-Reflective RAG (Self-RAG)~\cite{asai2023selfrag}} enhances an LM's quality and factuality through API retrieval and self-reflection.
(11) \textbf{Forward-Looking Active RAG (FLARE)~\cite{fiare}} iteratively uses a prediction of the upcoming sentence to anticipate future content and retrieve relevant documents to regenerate the sentence if it contains low-confidence tokens.
(12) \textbf{Dynamic Retrieval Augmented Generation based on the Information Needs (DRAGIN)~\cite{su2024dragin}} is designed to make decisions on when and what to retrieve based on real-time information needs.
Note that we strictly follow the prompts for the baselines as stated.

\subsubsection{Evaluation Metrics.}
To evaluate the performance of LLMs on multi-task medical choice questions, we instruct the models to provide only the correct answer and adopt the widely-used metric, \textbf{Exact Match (EM)}, as recommended by prior work~\cite{f1,f2}. An answer is deemed correct under the EM metric if its form exactly matches all the correct answers listed in the ground truth. The EM score is computed as follows:
\begin{equation*}
EM = \frac{\text{Number of Correct Answers}}{\text{Total Number of Answers}} \times 100\%.
\end{equation*}

For open-domain medical Q\&A tasks, we employ \textbf{ROUGE-R}~\cite{xu2023contextaware,HyKGE} and \textbf{Bilingual Evaluation Understudy (BLEU)} to assess the quality of the LLMs' responses. 

Specifically, for \textbf{BLEU}, \textbf{BLEU-1} is used to measure answer precision, and \textbf{BLEU-4} evaluates answer fluency by considering higher-order n-gram consistency.
\textbf{BLEU} evaluates the similarity of generated responses to the ground truth using the following formula:
\[
\text{BLEU-N} = BP \cdot \exp\left(\frac{1}{N} \sum_{n=1}^{N} \log p_n\right),
\]
where \( p_n \) is the precision of \( n \)-grams, \( BP \) is the Brevity Penalty, calculated as:
    \[
    BP = 
    \begin{cases} 
    1, & \text{if } c > r \\
    \exp\left(1 - \frac{r}{c}\right), & \text{if } c \leq r
    \end{cases}.
    \]
    Here \( c \) is the length of the generated response, and \( r \) is the length of the reference response.

\textbf{ROUGE-R} quantifies the recall of retrieved knowledge in the LLMs' responses, emphasizing their ability to comprehensively cover the information relevant to the query. For a generated response \( R \) and a reference \( G \), ROUGE-R is computed as:
\[
\text{ROUGE-R} = \frac{|R \cap G|}{|G|},
\]
where \( |R \cap G| \) denotes the number of overlapping n-grams between the generated response and the reference, and \( |G| \) is the total number of n-grams in the reference.

\subsubsection{Experimental Implementation.}
In \M, $\sigma=10$ for $cppl$ and 20 for $uct$. {Moreover, for all the baselines and \M, we set the maximum number of returned tokens for LLMs to 500 and the temperature to 0.6.} 
For a fair comparison, we apply the same W2NER, GTE~\cite{gte} models for all baselines. Moreover, the parameters of W2NER are optimized with Adam optimizer~\cite{Adam} with $L_2$ regularization and dropout on high-quality medical dataset~\cite{cmeee1,cmeee2}, the learning rate is set to 1e-3, the hidden unit is set to 1024 and weight decay is 1e-4. 
Implementations are done using the PyTorch 1.9.0 framework~\cite{paszke2019pytorch} in Python 3.9, on an Ubuntu server equipped with 8 A100 GPU and an Intel(R) Xeon(R) CPU.

\subsection{Additional Experiments}

\subsubsection{Generalization Experiments (RQ1)}
\label{appendix:Generalization Experiments}
To evaluate the generalization ability of \M, we conducted additional experiments on two MultihopQA datasets 2WikiMultihopQA~\cite{xanh2020_2wikimultihop} and HotpotQA~\cite{yang2018hotpotqadatasetdiverseexplainable} that require multihop reasoning, a reading comprehension dataset IIRC~\cite{fergusonirc} and a commonsense reasoning dataset StrategyQA~\cite{geva2021didaristotleuselaptop} in the general domain. All experiments utilized a unified Wiki-based knowledge repository as the RAG tool. Specifically, we employed the following LLMs: Llama3.2-1B-Instruct, LLaMA3.1-8B-Instruct~\cite{grattafiori2024llama3herdmodels}, Qwen2.5-7B-Instruct and Qwen2.5-14B-Instruct. 
For the two MultihopQA datasets and reading comprehension dataset, we use \textbf{Exact Match} at the answer level, along with token-level measurements of \textbf{F1-Score}, to assess the quality of responses. 
For the commonsense reasoning dataset, which consists of true-false questions, \textbf{EM} was employed as the primary evaluation metric.

\begin{table*}[t]
\caption{The general experimental results of \M~and other baselines on four benchmarks. The best results are in bold and the second runner are underlined.}
\label{tab:general}
\centering
\small
\begin{tabular}{ccccccccc}
\toprule
                                          &                 & \multicolumn{2}{c}{\textbf{2WikiMultihopQA}} & \multicolumn{2}{c}{\textbf{HotpotQA}} & \textbf{StrategyQA} & \multicolumn{2}{c}{\textbf{IIRC}} \\
                                          \midrule
\textbf{LLM}                              & \textbf{RAG Method} & \textbf{EM}          & \textbf{F1}           & \textbf{EM}       & \textbf{F1}       & \textbf{EM}         & \textbf{EM}     & \textbf{F1}     \\
\toprule
\multirow{6}{*}{\textbf{Llama3.2-1B-Instruct}}   
        & \textbf{wo-RAG} & 0.128 & 0.1958 & 0.067 & 0.1255 & 0.571 & 0.056 & 0.0728 \\
        & \textbf{SR-RAG} & {0.138} & {0.2052} & \underline{0.122} & \underline{0.2004} & {0.582} & \textbf{0.085} & \textbf{0.0922} \\
        & \textbf{FS-RAG} & \underline{0.143} & \underline{0.2094} & 0.120 & 0.1951 & \underline{0.589} & \underline{0.074} & \underline{0.0891} \\
        & \textbf{FL-RAG} & 0.134 & 0.1951 & 0.105 & 0.1892 & 0.548 & 0.072 & 0.0882 \\
        & \textbf{FLARE} & 0.136 & 0.1883 & 0.091 & 0.1575 & 0.572 & 0.050 & 0.0621 \\
        & \textbf{DRAGIN} & 0.126 & 0.1824 & 0.084 & 0.1334 & 0.570 & 0.051 & 0.0673 \\
        & \textbf{\M} & \textbf{0.153} & \textbf{0.2146} & \textbf{0.134} & \textbf{0.2164} & \textbf{0.591} & 0.066 & 0.0804 \\
                                          \toprule
\multirow{6}{*}{\textbf{Llama3.1-8B-Instruct}}       
        & \textbf{wo-RAG} & 0.260 & 0.3427 & 0.233 & 0.3223 & 0.747 & 0.190 & 0.2212 \\
        & \textbf{SR-RAG} & {0.339} & 0.4394 & 0.246 & 0.3550 & 0.726 & {0.221} & {0.2811} \\
        & \textbf{FS-RAG} & \underline{0.370} & \underline{0.4489} & \underline{0.316} & \underline{0.4298} & 0.745 & {0.218} & 0.2790 \\
        & \textbf{FL-RAG} & 0.336 & 0.4250 & {0.253} & 0.3618 & 0.722 & \underline{0.233} & \underline{0.2917} \\
        & \textbf{FLARE} & 0.286 & 0.3571 & 0.173 & 0.2710 & 0.745 & 0.197 & 0.2304 \\
        & \textbf{DRAGIN} & 0.260 & 0.3390 & 0.223 & 0.3167 & \underline{0.750} & 0.176 & 0.2168 \\
        & \textbf{\M} & \textbf{0.379} & \textbf{0.4654} & \textbf{0.340} & \textbf{0.4724} & \textbf{0.782} & \textbf{0.240} & \textbf{0.3132} \\
                                          \toprule
\multirow{6}{*}{\textbf{Qwen2.5-7B-Instruct}}  
        & \textbf{wo-RAG} & 0.141 & 0.2297 & 0.002 & 0.0662 & 0.583 & 0.068 & 0.1002 \\
        & \textbf{SR-RAG} & 0.157 & 0.2646 & 0.012 & 0.1039 & 0.581 & {0.156} & \underline{0.2084} \\
        & \textbf{FS-RAG} & \underline{0.222} & \underline{0.3284} & {0.047} & \textbf{0.1575} & {0.678} & \underline{0.158} & 0.2063 \\
        & \textbf{FL-RAG} & 0.182 & 0.2856 & 0.021 & 0.1360 & 0.628 & 0.138 & 0.1836 \\
        & \textbf{FLARE} & 0.126 & 0.2194 & 0.020 & 0.0970 & 0.618 & 0.095 & 0.1373 \\
        & \textbf{DRAGIN} & {0.195} & {0.2817} & \textbf{0.053} & 0.1339 & \underline{0.679} & 0.088 & 0.1241 \\
        & \textbf{\M} & \textbf{0.240} & \textbf{0.3478} & \underline{0.049} & \underline{0.1533} & \textbf{0.690} & \textbf{0.170} & \textbf{0.2307} \\
                                          \toprule
\multirow{6}{*}{\textbf{Qwen2.5-14B-Instruct}} 
        & \textbf{wo-RAG} & 0.037 & 0.1496 & 0.025 & 0.0595 & \textbf{0.780} & 0.048 & 0.0774 \\
        & \textbf{SR-RAG} & 0.073 & 0.1985 & 0.016 & 0.1310 & 0.731 & 0.105 & 0.1440 \\
        & \textbf{FS-RAG} & {0.148} & {0.2978} & \underline{0.117} & \underline{0.2412} & \underline{0.777} & \textbf{0.213} & 0.2556 \\
        & \textbf{FL-RAG} & 0.080 & 0.2347 & 0.040 & 0.1702 & \textbf{0.780} & 0.140 & \underline{0.2580} \\
        & \textbf{FLARE} & \underline{0.254} & \underline{0.3621} & 0.025 & 0.1208 & 0.721 & 0.071 & 0.1019 \\
        & \textbf{DRAGIN} & 0.220 & 0.3231 & 0.100 & 0.2168 & 0.700 & 0.120 & 0.1624 \\
        & \textbf{\M~} & \textbf{0.370} & \textbf{0.4397} & \textbf{0.180} & \textbf{0.2604} & \textbf{0.780} & \underline{0.200} & \textbf{0.2915} \\
                                          \toprule
\end{tabular}
\end{table*}

\subsubsection{Noise Poisoning Attack (RQ3)}
\label{Noise Poisoning Attack}
To assess the robustness of \M~against noise poisoning attacks, we conducted two distinct types of attacks: \textbf{Partial Attack and Structural Attack}. These attacks simulate scenarios involving excessive redundant information or situations where retrieval mechanisms fail, leaving no effective information. Additionally, we introduced two categories of noise—\textbf{Irrelevant Retrieval Noise \& Relevant Retrieval Noise}~\cite{yoran2024makingretrievalaugmentedlanguagemodels,cuconasu2024power,fang2024enhancingnoiserobustnessretrievalaugmented,zhang2024knowpoknowledgeawarepreferenceoptimization}—to evaluate the effectiveness of the \texttt{Backtrack} and \texttt{Summary} actions.

In terms of implementation, we artificially constructed 100 pieces of noise based on CMB-Exam for each type of noise. It is worth noting that for partial attacks, we directly concatenate the noise knowledge after the retrieved results. For structural attacks, we replace the retrieved knowledge instead. We conduct the poison attack during the first function call:

\begin{table}[h]
\centering
\caption{Results for Structural and Partial Attacks}
\label{table: attack}
\resizebox{\linewidth}{!}
{
\begin{tabular}{lccc}
\toprule
\textbf{Noise Type} & \textbf{\texttt{Summary} Prob.} & \textbf{\texttt{Backtrack} Prob.} & \textbf{EM} \\
\midrule
\multicolumn{4}{l}{\textbf{Structural Attack}} \\
\midrule
Irrelevant Retrieval Noise & 01.00\% & 94.00\% & 91.00 \\
Relevant Retrieval Noise   & 12.00\% & 76.00\% & 84.00 \\
Mixed Retrieval Noise      & 04.00\% & 91.00\% & 89.00 \\
\midrule
\multicolumn{4}{l}{\textbf{Partial Attack}} \\
\midrule
Irrelevant Retrieval Noise & 37.00\% & 49.00\% & 85.00 \\
Relevant Retrieval Noise   & 52.00\% & 29.00\% & 79.00 \\
Mixed Retrieval Noise      & 41.00\% & 47.00\% & 81.00 \\
\bottomrule
\end{tabular}}
\end{table}

\begin{table*}[ht]
\centering
\caption{Performance of Various Approaches on CMB Leaderboard}
\resizebox{\textwidth}{!}{
\begin{tabular}{l|c|cccccc}
\hline
\multirow{2}{*}{\textbf{Method}} & \textbf{Average} & \textbf{Medical} & \textbf{Nursing} & \textbf{Pharmacist} & \textbf{Medical} & \textbf{Professional} & \textbf{Medical} \\ 
& \textbf{Score} & \textbf{Qualification} & \textbf{Exam} & \textbf{Exam} & \textbf{Technician} & \textbf{Knowledge} & \textbf{Postgraduate} \\ \hline
\textbf{\M}      & \textbf{87.95}                  & \textbf{89.50}                              & \textbf{92.63}                 & \textbf{88.00}                    & \textbf{86.75}                           & \textbf{83.56}                                 & \textbf{87.25}                               \\ \hline
HuatuoGPTII-34B     & 76.80                  & 75.65                              & 82.31                 & 76.81                    & 76.17                           & 74.38                                 & 75.56                               \\ \hline
Qwen-72B-Chat       & 74.38                  & 78.55                              & 83.56                 & 79.78                    & 77.92                           & 68.25                                 & 58.19                               \\ \hline
Yi-34B-Chat         & 69.17                  & 71.10                              & 77.56                 & 73.16                    & 73.67                           & 66.56                                 & 52.94                               \\ \hline
Yi-6B-Chat          & 65.87                  & 67.25                              & 76.38                 & 68.50                    & 67.83                           & 61.75                                 & 53.50                               \\ \hline
GPT-4               & 59.46                  & 59.90                              & 69.31                 & 52.19                    & 61.50                           & 59.69                                 & 54.19                               \\ \hline
Qwen-14B-Chat       & 57.64                  & 60.40                              & 65.63                 & 60.94                    & 58.83                           & 54.50                                 & 45.56                               \\ \hline
Baichuan2-13B-chat  & 39.88                  & 40.04                              & 45.65                 & 40.60                    & 39.25                           & 39.25                                 & 34.45                               \\ \hline
ChatGLM2-6B         & 38.51                  & 40.25                              & 47.56                 & 36.06                    & 36.58                           & 35.56                                 & 35.06                               \\ \hline
Baichuan-13B-chat   & 38.20                  & 37.70                              & 44.75                 & 41.22                    & 34.67                           & 37.94                                 & 32.94                               \\ \hline
HuatuoGPT           & 29.49                  & 29.90                              & 34.00                 & 29.06                    & 30.92                           & 27.38                                 & 25.69                               \\ \hline
ChatMed-Consult     & 20.23                  & 19.40                              & 21.69                 & 20.00                    & 22.83                           & 18.88                                 & 18.56                               \\ \hline
\end{tabular}
}
\label{tab:cmb model_performance}
\end{table*}

\paragraph{Impact of Structural Attack and Partial Attack}
In the context of \textbf{Structural Attack}, the influence of \textbf{Partial Attack} is particularly pronounced. Since this type of noise is directly embedded within the sentence, it exerts a greater interference on \M. The experimental results indicate that under Partial Attack, (1) \M~is more likely to trigger the \texttt{Summary} action, attempting to condense and process the excess information. This suggests that the model tends to utilize summarization as a means to handle noise in such scenarios.
(2) Nevertheless, \M~still triggers the \texttt{Backtrack} action in some cases, though with a lower probability compared to \textbf{Structural Attack}.
(3) More notably, due to the greater impact of this embedded noise, the model's EM score significantly decreases, indicating that \textbf{Partial Missing} noise has the most substantial impact on \M~during \textbf{Structural Attack}.

\paragraph{Comparison of Noise Types}
(1) In contrast, the impact of \textbf{Relevant Noise} is even more severe, particularly in the context of \textbf{Partial Attack}. Since \textbf{Relevant Noise} is highly related to the task, \M~struggles to determine whether the noise contains the required answer, leading to a significantly lower EM compared to when dealing with \textbf{Irrelevant Noise}.
(2) \textbf{Relevant Noise} is more likely to trigger the \texttt{Summary} action, indicating that when faced with task-related noise, the model may prefer summarizing the information rather than directly identifying and discarding irrelevant content.
(3) In contrast, \textbf{Irrelevant Noise} is more easily detected by \M~and effectively removed through the \texttt{Backtrack} action. The model handles it more efficiently, with the \texttt{Backtrack} execution probability reaching as high as 94\%. 
(4) The results for \textbf{Mixed Noise} fall between the two, but since it contains \textbf{Irrelevant Noise}, which is easier for the model to detect, its performance is closer to that of \textbf{Irrelevant Noise}.

\paragraph{Overall Robustness under Attack}
Overall, under the attacks, \M~demonstrates strong robustness, with the execution probabilities of \texttt{Summary} and \texttt{Backtrack} remaining above 81\%, and sometimes reaching as high as 95\%. This clearly illustrates the effectiveness of \M~in managing the memory stack, effectively preventing the introduction of erroneous knowledge and irrelevant information, thereby maintaining the purity of the memory stack, in line with \textbf{\textit{C3}}.

\subsubsection{Case Study (RQ3) }
\label{Case Study}
In this case study, we examine the effectiveness of two RAG  systems: one without memory and state management--ReACT, and \M. We evaluate how each system impacts the state management, retrieval, and reasoning process, particularly when dealing with irrelevant or incorrect noise.

\begin{figure*}[h]
   \centering
 \includegraphics[width=\textwidth]{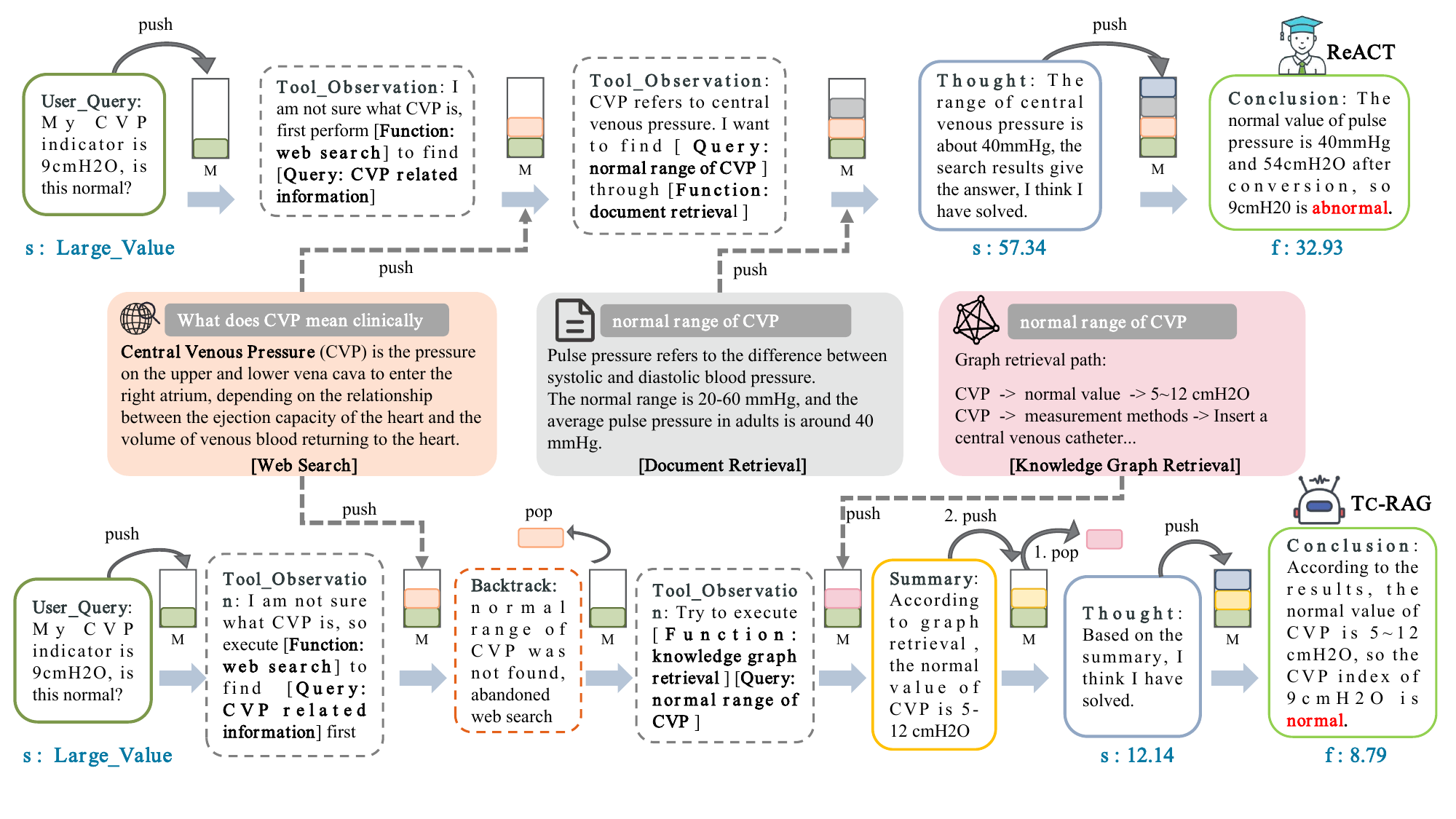}
   \caption{Case Study of ReACT-based and \M.}
   \label{fig:case.png}
 \end{figure*}

The results indicate that the ReACT-based approach struggles with accumulating irrelevant noise, leading to overconfidence and inaccurate conclusions, simply due to the unit conversion in Figure~\ref{fig:case.png}.
In contrast, \M~effectively manages its memory and utilizes \texttt{Summary} \& \texttt{Backtrack} actions to prune incorrect retrievals, resulting in more concise and accurate conclusions, which underscores \M's superiority in handling complex tasks (for \textbf{\textit{C3}}).
Furthermore, we found that the ReACT-based approach tends to prematurely settle on answers when the system state value is high, due to the lack of state management. On the other hand, \M~dynamically monitors the RAG process, ensuring that the system state value meets the termination condition, which highlights the necessity of constructing a system state, in line with \textbf{\textit{C1}}.

\subsubsection{In-Depth Analysis of the Trajectories of \M~ with Baselines}

We expand the analysis with an additional case study on the user query:

\textbf{User Query:} “My CVP indicator is 9 cmH2O, is this normal?”

For this, we provide and compare the complete reasoning trajectories of multiple baselines, including Base, CoT, FLARE, and DRAGIN, in addition to TC–RAG.

\textbf{Base:} \textit{Answer:} "normal" is True, but the CVP normal range is False.

Central Venous Pressure (CVP) is a measure of the pressure in the thoracic vena cava, near the right atrium of the heart. It is an important indicator of fluid status and right heart function. The normal range for CVP is typically between 2 to 6 cmH2O, although some sources may extend this range slightly to 0 to 8 cmH2O. A CVP of 9 cmH2O is slightly above the typical normal range.

\textbf{FLARE:} \textit{Answer:} "normal" is True, the CVP normal range is True.

\begin{quote}
Sentence 1: Central Venous Pressure (CVP) is a measure of the pressure in the thoracic vena cava, near the right atrium of the heart. (search not triggered)

Sentence 2: It is an important indicator of fluid status and right heart function. The \textbf{normal range for CVP} is typically between 2 to 6 cmH2O. (search triggered)

Search: normal range for CVP

Observation: normal range of CVP is between 5-12 cmH2O.

Sentence 2: It is an important indicator of fluid status and right heart function. The normal range for CVP is typically between 5 to 12 cmH2O. (search not triggered)

Sentence 3: Therefore, having a CVP of 9 cmH2O is normal. (search not triggered, ended)
\end{quote}

\textbf{DRAGIN:} \textit{Answer:} "normal" is True, the CVP normal range is True.

\begin{quote}
Sentence 1: Central Venous Pressure (CVP) is a measure of the pressure in the thoracic vena cava, near the right atrium of the heart. (search not triggered)

Sentence 2: It is an important indicator of fluid status and right heart function. The \textbf{normal range for CVP} is typically between 2 to 6 cmH2O. (search triggered)

Search: normal range for Central Venous Pressure

Observation: normal range of CVP is between 5-12 cmH2O.

Sentence 2: Therefore, having a CVP of 9 cmH2O is normal. (search not triggered, ended)
\end{quote}

\textbf{ReACT:} \textit{Answer:} "normal" is False, the CVP normal range is False.

\begin{quote}
Step 1: Tool Observation: I am not sure what CVP is. First perform [Function: web search] to find [Query: CVP related information].

Step 2: Tool Observation: CVP refers to central venous pressure. I want to find [Query: normal range of CVP] through [Function: document retrieval].

Step 3: Thought: The range of central venous pressure is about 40mmHg, the search results give the answer, I think I have solved.

Step 4: Conclusion: The normal value of pulse pressure is 40mmHg and 54cmH2O after conversion, so 9cmH2O is abnormal.
\end{quote}

\textbf{\M~:} \textit{Answer:} "normal" is True, the CVP normal range is True.

\begin{quote}
Step 1 -- push: Tool Observation: I am not sure what CVP is, so execute [Function: web search] to find [Query: CVP related information] first.

Step 2 -- pop: Backtrack: normal range of CVP was not found, abandoned web search.

Step 3 -- push: Tool Observation: Try to execute [Function: knowledge graph retrieval] [Query: normal range of CVP].

Step 4 -- pop \& push: Summary: According to graph retrieval, the normal value of CVP is 5-12 cmH2O.

Step 5 -- push: Thought: Based on the summary, I think I have solved.

Step 6 -- push: Conclusion: According to the results, the normal value of CVP is 5-12 cmH2O, so the CVP index of 9cmH2O is normal.
\end{quote}

The comparative results reveal that:

\begin{itemize}
    \item The ReACT approach tends to accumulate irrelevant information across steps, leading to a noisy context, inflated confidence, and ultimately incorrect conclusions. In the original case, even a simple unit mismatch introduces confusion that remains uncorrected due to the lack of memory revision mechanisms.
    \item FLARE and DRAGIN demonstrate adaptive information retrieval capabilities, which reduces the probability of acquiring noisy information when LLM's internal knowledge suffices to solve the problem. However, if irrelevant or harmful information is inadvertently included in the memory, these models cannot eliminate the impact of such noise.
    \item In contrast, TC–RAG effectively leverages its stack-based memory, employing Summary and Backtrack actions to eliminate erroneous information and maintain a coherent and accurate reasoning trace, resulting in a more precise conclusion.
\end{itemize}

\subsubsection{The Ability of Backtracking}

\paragraph*{1. Noise Prevention Through Pre-Insertion Reasoning}

\M~ is based on a simple yet practical assumption: if each piece of knowledge is carefully evaluated before being added to memory, the risk of accumulating noisy or irrelevant information deep in the stack is substantially reduced. This preemptive filtering ensures that only relevant and reliable information is retained, meaning any potential noise typically appears near the top of the stack. As a result, local backtracking is generally sufficient in practice. This assumption underlies our choice of a stack-based memory structure, which enables efficient \(O(1)\) access to recent reasoning steps, thereby optimizing both accuracy and computational efficiency.

\paragraph*{2. Maintaining Reasoning Consistency Through Sequential Memory}

Allowing arbitrary modifications to deep memory entries would disrupt the sequential and causal continuity of the reasoning process. As discussed in Section 5 (Lines 433–437), once information is committed to the memory stack, it becomes part of the model’s reasoning trajectory, where each new memory is conditioned on the previous ones. Direct edits to earlier entries may introduce inconsistencies or logical contradictions, particularly in domains that demand high levels of traceability and safety. Therefore, \M~ deliberately restricts memory modifications to the top of the stack in order to preserve the integrity, coherence, and interpretability of the reasoning flow.

\paragraph*{3. Iterative Backtracking as a Practical Fallback}

In cases where outdated or incorrect information is later found to reside deeper in the stack, \M~ supports progressive correction through (1) multiple successive backtracking steps, or (2) higher-level action, such as defining a root-cause tracing action, that internally performs a sequence of pop (\(>1\)) operations. This mechanism enables the system to rewind its reasoning trajectory larger when necessary, without compromising the temporal structure of the memory or the continuity of the decision process.

Here, we do also add a real case example from our log in the CMB dataset in which pop action appears more than once.

\begin{quote}
\textbf{Query (push):} A patient with a history of type 2 diabetes for 8 years is taking Metformin 1000mg bid and Dapagliflozin 10mg qd. Recently, he has experienced worsening fatigue, frequent nocturia, and foot numbness. The eGFR is 58 ml/min. What's the problem?\\
\textbf{Thought (push):} The patient's symptoms may be related to diabetes. Let me check it.\\
\textbf{Search (push):} Symptoms of diabetes\\
\textbf{Observation (push):} Common symptoms of diabetes include frequent urination, excessive thirst, unexplained weight loss, fatigue... (omitted)\\
=====> \textbf{[Backtrack (pop):]} diabetes symptoms information is not helpful.\\
=====> \textbf{[Backtrack (pop):]} I should change my search question.\\
\textbf{Thought (push):} The medicine may be relevant to the patient's symptoms.\\
\textbf{Search (push):} metformin and dapagliflozin\\
\textbf{Observation (push):} ... When eGFR < 60ml/min, Dapagliflozin needs to be reduced to 5mg qd ... (omitted)\\
\textbf{Thought (push):} The eGFR of the patient is 58ml/min. I think I found the reason.\\
\textbf{Final Answer (push):} The patient's eGFR suggests possible renal injury, and the symptoms are due to excessive Dapagliflozin dosage, requiring dose reduction.
\end{quote}

It can be observed that when an unrelated message resides deeper within the memory stack, the model will expunge irrelevant memories via sequential popping operations upon detection.

In summary, \M~ adopts a cautious and principled approach that balances efficiency, interpretability, and reasoning stability. 

\subsection{How Does the System Handle Instructions Failing or Formatting Errors}
Similar to other multi-step reasoning agent frameworks such as ReAct~\cite{react}, LangChain, LangGraph, and AutoGPT~\cite{autogpt}, the probabilistic nature of LLMs can occasionally lead to outputs that deviate from the specified format or fail to fully comply with instructions. We refer to the methods in~\cite{react,autogpt} for handling exceptions:

\begin{enumerate}[leftmargin=*]
    \item When the LLM generates an output that does not strictly adhere to the required format, we first employ string manipulation techniques (e.g., regex matching, template-based extraction) to identify and isolate valid components within the response. This allows us to salvage usable information even when the overall structure is imperfect.
    
    \item If parsing still fails after multiple attempts, we leverage a retry mechanism to improve the likelihood of generating a parsable response. The system uses a while loop and try-catch blocks to force the LLM to re-generate output in the required format. Only when it generates parsable output or reaches a predefined maximum limit can the system stop.
\end{enumerate}

\subsection{CMB Leaderboard}
\label{Leaderboard}
Here, we compared the specific testing results of our model on CMB with the publicly available rankings on CMB Leaderboard with open-source baselines in Table~\ref{tab:cmb model_performance}. The experimental results show that \M~has surpassed baselines in all medical indicators.